\documentclass{aa}
\usepackage{graphicx}
\usepackage{bm,natbib,url,wasysym,color}
\usepackage[varg]{txfonts}
\bibpunct{(}{)}{;}{a}{}{,} 
\graphicspath{{./figs/}{./png/}}

\newcommand{\brac}[1]{\langle #1 \rangle}
\newcommand{\EQ}{\begin{equation}}
\newcommand{\EN}{\end{equation}}
\newcommand{\EQA}{\begin{eqnarray}}
\newcommand{\ENA}{\end{eqnarray}}

\newcommand{\Eq}[1]{Equation~(\ref{#1})}

\newcommand{\Eqss}[2]{Equations~(\ref{#1})--(\ref{#2})}

\newcommand{\App}[1]{Appendix~\ref{#1}}
\newcommand{\Sec}[1]{Section~\ref{#1}}
\newcommand{\Secs}[2]{Sections~\ref{#1} and \ref{#2}}
\newcommand{\Fig}[1]{Fig.~\ref{#1}}

\newcommand{\Figs}[2]{Figs.~\ref{#1} and \ref{#2}}

\newcommand{\Tab}[1]{Table~\ref{#1}}

\newcommand{\fluc}[1]{#1^\prime}
\newcommand{\mean}[1]{\overline #1}
\newcommand{\meanrho}{\overline{\rho}}

{}
{}
{}

\newcommand{\meanuu}{\overline{\mbox{\boldmath $u$}}{}}{}
{}
{}
{}
{}
{}
{}
{}
{}
\newcommand{\meanBB}{\overline{\mbox{\boldmath $B$}}{}}{}
{}
{}
{}
{}
{}
{}
{}
{}
\newcommand{\meanUU}{\overline{\bm{U}}}

{}
{}
{}

{}

{}
{}

\newcommand{\Ot}{\tilde{\Omega}}

\newcommand{\eee}{\hat{\mbox{\boldmath $e$}} {}}

\newcommand{\gggg}{\mbox{\boldmath $g$} {}}

\newcommand{\rr}{\mbox{\boldmath $r$} {}}

\newcommand{\uu}{\mbox{\boldmath $u$} {}}

\newcommand{\ssss}{\mbox{\boldmath $\xi$} {}}

\newcommand{\BB}{\mbox{\boldmath $B$} {}}

\newcommand{\JJ}{\mbox{\boldmath $J$} {}}

\newcommand{\AAA}{\mbox{\boldmath $A$} {}}

\newcommand{\FF}{\mbox{\boldmath $F$} {}}

\newcommand{\nab}{\mbox{\boldmath $\nabla$} {}}
\newcommand{\OO}{\bm{\Omega}}
\newcommand{\oo}{\mbox{\boldmath $\omega$} {}}

\newcommand{\SSSS}{\mbox{\boldmath ${\sf S}$} {}}

\newcommand{\DD}{{\rm D} {}}

\newcommand{\dd}{{\rm d} {}}

\def\degr{\hbox{$^\circ$}}

\def\Ta{\mbox{\rm Ta}}
\def\Ra{\mbox{\rm Ra}}

\def\Co{\mbox{\rm Co}}

\def\Pr{\mbox{\rm Pr}}

\def\Pm{\mbox{\rm Pr}_{\rm M}}
\def\PrS{\mbox{\rm Pr}_{\rm SGS}}
\def\Rm{\mbox{\rm Re}_{\rm M}}

\def\Rey{\mbox{\rm Re}}
\newcommand{\Rc}{{{R_{C}}}}
\def\Pe{\mbox{\rm Pe}}
\def\Co{\mbox{\rm Co}}

\def\kf{k_{\rm f}}

\def\urms{u_{\rm rms}}
\def\urmsp{u^\prime_{\rm rms}}

\def\nut{\nu_{\rm t}}

\def\Btt{\overline{B}{}_\phi^{\rm rms}}

\def\half{{\textstyle{1\over2}}}

\def\onethird{{\textstyle{1\over3}}}

\newcommand{\yr}{\,{\rm yr}}

\newcommand{\chiS}{\chi_{\rm SGS}}
\newcommand{\chiSm}{\chi_{\rm SGS}^{m}}
\newcommand{\LLM}{\Lambda_{\rm M}}
\newcommand{\LLH}{\Lambda_{\rm H}}
\newcommand{\LLV}{\Lambda_{\rm V}}
\begin{document}

\titlerunning{Influence of a coronal envelope on global simulations}
\authorrunning{Warnecke et al.}

\title{Influence of a coronal envelope as a free boundary\\ to global convective
  dynamo simulations}
\author{J. Warnecke\inst{1,2}
\and P. J.\ K\"apyl\"a\inst{3,2,1,4} 
\and M. J.\  K\"apyl\"a\inst{1,2}
\and A. Brandenburg\inst{4,5,6,7}}
\institute{Max-Planck-Institut für Sonnensystemforschung,
  Justus-von-Liebig-Weg 3, D-37077 G\"ottingen, Germany\\
\email{warnecke@mps.mpg.de}\label{inst1} 
\and ReSoLVE Centre of Excellence, Department of Computer Science,
Aalto University, PO Box 15400, FI-00076 Aalto, Finland \label{inst2}
\and Leibniz-Institut f\"ur Astrophysik Potsdam, 
An der Sternwarte 16, D-11482 Potsdam, Germany \label{inst3}
\and NORDITA, KTH Royal Institute of Technology and Stockholm University,
Roslagstullsbacken 23, SE-10691 Stockholm, Sweden\label{inst4}
\and Department of Astronomy, AlbaNova University Center, Stockholm
University, SE-10691 Stockholm, Sweden\label{inst5}
\and JILA and Department of Astrophysical and Planetary Sciences, Box
440, University of Colorado, Boulder, CO 80303, USA\label{inst6}
\and Laboratory for Atmospheric and Space Physics, 3665 Discovery
Drive, Boulder, CO 80303, USA\label{inst7}}
\date{Received 18 March 2015 / Accepted 21 July 2016}
\abstract{}{%
We explore the effects of an outer stably stratified coronal envelope on rotating
turbulent convection, differential rotation, and large-scale dynamo action
in spherical wedge models of the Sun.
}{%
We solve the compressible magnetohydrodynamic equations in a two-layer
model with unstable stratification below the surface,
representing the convection zone, and a stably
stratified coronal envelope above.
The interface represents a free surface.
We compare our model to models that have no coronal envelope.
}{%
The presence of a coronal envelope is found to modify the Reynolds
stress and the $\Lambda$ effect resulting in a weaker and
non-cylindrical differential rotation.
This is related to the reduced latitudinal temperature variations that
are caused by and dependent on the angular velocity.
Some simulations develop a near-surface shear
layer that we can relate to a sign change in the meridional
Reynolds stress term in the thermal wind balance equation.
Furthermore, the presence of a free surface changes the
magnetic field evolution since the toroidal field is concentrated
closer to the surface.
In all simulations, however, the migration direction of the mean magnetic
field can be explained by the Parker--Yoshimura rule,
which is consistent with earlier findings.
}{%
A realistic treatment of the upper boundary in spherical dynamo simulations
is crucial for the dynamics of the flow and magnetic field evolution.
}
\keywords{Magnetohydrodynamics (MHD) -- turbulence -- dynamo -- Sun:
  magnetic fields -- Sun: rotation-- Sun: activity
}

\maketitle

\section{Introduction}
The Sun has an activity cycle of about 11 years, with an underlying
magnetic field that oscillates with a period of around 22 years.
A dynamo operating in the convection zone below the solar surface is
responsible for generating cyclic magnetic fields
\citep[see e.g.,][and reference therein]{BS05,Char05}.
The occurrence of sunspots, which is the main surface manifestation
of the solar cycle, varies regularly over the cycle.
At the beginning of each cycle sunspots tend to appear at mid-latitudes,
while toward the end of each cycle they tend to appear at low latitudes.
It is believed that these sunspots and their occurrence are connected
to an underlying toroidal magnetic field, which is migrating
equatorward during the cycle.
Theoretical models of solar magnetic field evolution have been studied
for several decades.
Mean-field models, where turbulence effects are
parameterized through transport coefficients \citep[see e.g.,][]{KR80}, 
have been successful in producing some observed magnetic field
properties \citep[e.g.,][]{KKT06,KO11}. 
Another class of dynamo models
relies on the Babcock-Leighton effect \citep{B61,L64} and flux
transport by meridional circulation \citep[e.g.,][]{CSD95,DC99}.
The low Reynolds numbers compared with the Sun limit the usefulness of
global simulations of self-consistent convection, where the
equations of magnetohydrodynamics are solved directly.
However, the increasing computing power has led to the successful
reproduction of some observed features of the solar magnetic field by
such models.

For a long time global simulations were only able to generate either
poleward migrating fields \citep{G83,BMT04,KKBMT10,NBBMT13} or
oscillatory ones with no clear migration pattern \citep{GCS10,Racine11}.
For the first time, \cite{SPD11} and \cite{KMB12} could produce clear
equatorward migration of the toroidal magnetic field.
Stratification and rotation rate had to be high enough for this
to work \citep{KMCWB13}.
Recently, several groups have been able to produce grand minima-type
events \citep{PC14,ABMT15,KKOBWKP16}.
In \cite{KKOBWKP16}, a secondary dynamo mode disturbed the surface
field of the primary mode.
\cite{WKKB14} could explain the equatorward migration seen in
simulations of \cite{KMB12,KMCWB13} and those of \cite{ABMT15} as a
propagating $\alpha\Omega$ dynamo wave following the Parker--Yoshimura
rule \citep{P55,Yos75}.
Here, $\alpha$ was estimated via the kinetic helicity and $\Omega$ is
the local solar rotation rate.
An equatorward migrating dynamo wave is possible if $\alpha$ is
positive (negative) in the northern (southern) hemisphere
\citep{SKR66} and the radial gradient of $\Omega$ is negative.
These interpretations have been confirmed independently by \cite{DWBG16}
and by the computation of turbulent transport coefficients obtained through the test-field method
\cite{WRKKB16}.

Given that this simple relation can describe the behavior of dynamos
driven by self-consistent turbulent convection in simulations, the
Parker--Yoshimura rule can also be a possibility to
explaining the equatorward migration of the magnetic field of the Sun.
Indeed, the differential rotation of the Sun \citep{Schouea98} has a
negative radial gradient in the near-surface shear layer
\citep{Thompson96,BSG14,BSG16}.
This makes it a possible location of the solar dynamo \citep{B05}.

The generation of the solar differential rotation can well be
described by mean-field models, where the productive parts of the
off-diagonal Reynolds stress are parameterized by the $\Lambda$ effect
\citep{R89} and the turbulent heat transport in terms of anisotropic
turbulent heat conductivity \cite[see also][]{BMT92b}.
These models can reproduce the surface differential rotation, the
spoke-like rotation profile and the
near-surface shear layer \citep[e.g.,][]{KR95,KR05,RKH13}.
With global models of turbulent convection, it is challenging to generate
such rotation profiles \citep[e.g.,][]{DGT02,B07}.
\cite{MBT06} were able to produce a spoke-like rotation profile by
imposing a latitudinal entropy gradient at the bottom boundary, while
\cite{BMT11} used a
stably stratified layer below the convection zone.
\cite{GSKM13} could produce a near-surface shear layer at lower
latitudes.
Recently, \cite{HRY15} used a reduced sound speed technique \citep{HRYIF12}
achieving high stratification to produce a near-surface shear layer,
the generation of which was suggested to be related with the meridional Reynolds
stress component $Q_{r\theta}$. 
However, they still had cylindrical
rotation contours within the convection zone.

\cite{WKMB13} applied a different approach and used an outer coronal
envelope above the dynamo domain \citep{WB10} to reproduce
spoke-like differential rotation at low latitudes with
a weak near-surface shear layer.
This two-layer approach has also been used to successfully simulate
coronal ejections driven by dynamos arising from forced turbulence
\citep{WBM11,WBM12}
as well as by convective dynamos \citep{WKMB12}.
Furthermore, the outcome of dynamo simulations suggests that the
presence of a coronal envelope supports the dynamo and leads to a
higher field strength \citep{WB14}.

To investigate and follow up on the findings of \cite{WKMB13}, we perform a
detailed study of similar simulations with and without a coronal envelope to
investigate the effect of a coronal envelope as a free boundary on a
convectively driven dynamo.
We vary the size of the
envelope, as well as the cooling profile, the magnetic boundary
condition, and the rotation rate.
We analyze the effect on the flows, differential rotation, and the
magnetic field evolution.
Even though the solar corona most likely has limited influence on
the dynamics of subsurface flows and the evolution of the magnetic
fields in the Sun, these studies are important for investigating different
influences and effects on convective dynamo simulations.
Every simulation, in which we better understand the mechanism causing flow and
magnetic field evolution, will bring us a step closer toward understanding
the dynamics of the interior of the Sun and other stars.

\section{Model and setup}
\label{sec:model}

\begin{figure}[t!]
\begin{center}
\includegraphics[width=0.9\columnwidth]{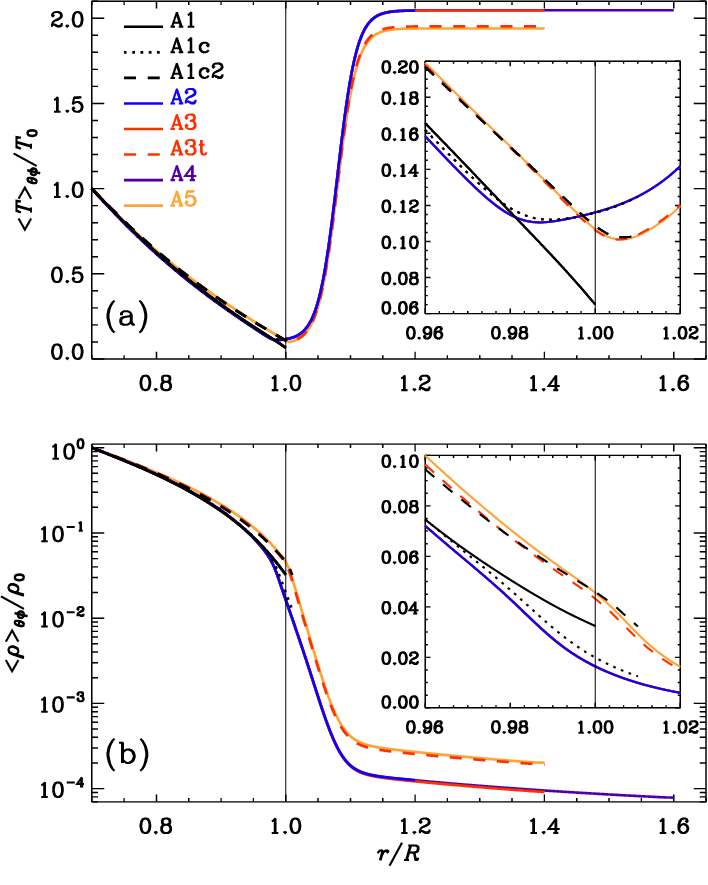}
\end{center}\caption[]{
Radial profiles of azimuthally and latitudinally averaged temperature
$\brac{T}_{\theta\phi}$ (a) and density $\brac{\rho}_{\theta\phi}$ (b)
normalized by their values at the bottom of the domain, $T_0$ or
$\rho_0$, respectively for Set~A.
The inlays show the radial profiles near the surface.
}
\label{strat1_A}
\end{figure}

Our setup is similar to the one-layer model of
\cite{KMB12,KMCWB13} and the two-layer model of \cite{WKMB13},
both of which have recently also been used in \cite{WKKB14}.
We use a wedge in spherical polar coordinates ($r,\theta,\phi$), in which
the layer below the surface ($r_0\le r\le R$) represents the convection zone.
Here, $R$ is the solar radius and $r_0$ corresponds to the bottom of the
convection zone at $r=0.7\, R$.
The layer above the surface ($R\le r\le\Rc$) represents a simplified
coronal envelope, which extends to outer radius $\Rc$.
The domain spans $15^\circ\le\theta\le 165^\circ$ in colatitude
and $0^\circ\le\phi\le90^\circ$ in longitude.
We solve the equations of compressible magnetohydrodynamics,
\begin{equation}
{\partial\AAA\over\partial
  t}=\uu\times\BB-\mu_0\eta\JJ,
\end{equation}
\begin{equation}
{\DD\ln\rho\over\DD t} =-\nab\cdot\uu,
\end{equation}
\begin{equation}
{\DD\uu\over\DD t}=  \gggg - 2\OO_0 \times \uu + {1\over\rho}
\left(\JJ\times\BB - \nab p+\nab\cdot 2\nu\rho\SSSS\right),
\end{equation}
\begin{equation}
T{\DD s\over\DD t}=-{1\over\rho}\nab\cdot
\left(\FF^{\rm rad}+ \FF^{\rm SGS}\right) +
2\nu\SSSS^2+{\mu_0\eta\over\rho}\JJ^2 -
\Gamma_{\rm cool},
\label{entro}
\end{equation}
where the magnetic field is defined via the vector potential
$\BB=\nab\times\AAA$, ensuring the solenoidality of $\BB$ at all
times, $\JJ=\nab\times\BB/\mu_0$ is the current density with $\mu_0$
being the vacuum permeability, $\eta$ is the magnetic diffusivity,
$\nu$ is the kinematic viscosity,
$\uu$ is the plasma velocity, $\rho$ is the mass density, $s$ is the
specific entropy, and
$\DD/\DD t=\partial/\partial t + \uu\cdot\nab$ is the
Lagrangian derivative.
The traceless rate-of-strain tensor is given by
\begin{equation}
{\mathsf
  S}_{ij}=\half(u_{i;j}+u_{j;i})-\onethird\delta_{ij}\nab\cdot\uu,
\end{equation}
where semicolons denote covariant differentiation; see \cite{MTBM09}
for details.
The gravitational acceleration is given by
\begin{equation}
\gggg=-GM\rr/r^3,
\end{equation}
where $G$ is Newton's gravitational constant and $M$ is the
mass of the Sun.
In addition $\OO_0=\Omega_0 (\cos\theta,-\sin\theta,0)$ is the
rotation vector,
where $\Omega_0$ is the rotation rate of the comoving frame.
Using the ideal gas law, the pressure is given by $p=(\gamma-1)\rho e$,
where $\gamma=c_{\rm P}/c_{\rm V}=5/3$ is the ratio of specific heats at constant
pressure and constant volume, respectively, and $e=c_{\rm V} T$ is the
internal energy density, which is related to the temperature $T$.
The two diffusive heat fluxes are defined as 
\begin{equation}
\FF^{\rm rad}=-K\nab T,\quad \FF^{\rm SGS} =-\chiS \rho  T\nab s,
\end{equation}
where $\FF^{\rm rad}$ is the radiative heat flux with the radiative
heat conductivity $K$ and $\FF^{\rm SGS}$ is the subgrid scale
(SGS) heat flux that carries the unresolved turbulent heat flux of
convection with the SGS heat diffusivity $\chiS$; see
\cite{KMCWB13} and \cite{WKMB13} for details.
Finally, the function $\Gamma_{\rm cool}$ relaxes the temperature toward
a predefined profile $T_{\rm ref}(r)$
\begin{equation}
\Gamma_{\rm cool}=\Gamma_0 f(r){T-T_{\rm ref}(r)\over T_{\rm ref}(r)},
\label{cool}
\end{equation}
where $\Gamma_0$ is a cooling luminosity. $f(r)$ is a profile function
tending to unity in $r>R$
and going smoothly to zero in $r\leq R$,
see \cite{WKMB13} for details. Figure~\ref{strat1_A} shows the corresponding
temperature and density stratifications for the runs in Set~A,
see \Tab{runs}.

We use isentropic, hydrostatic initial conditions, as in previous models
\citep{KMCWB13,WKMB13}.
This initial setup is not in thermal equilibrium, but the flux at
the lower boundary exceeds the flux leaving at the outer boundary,
resulting the onset of convective instability.
Furthermore, we initialize the magnetic field as a
white noise seed field in the convection zone.
We apply periodic boundary conditions in the azimuthal ($\phi$)
direction.
For the velocity field we apply stress-free boundary conditions at the
radial and latitudinal boundaries.
The magnetic field follows a perfect conductor condition at the lower
radial and at the latitudinal boundaries.
We force the field to be radial at the top boundary.
Furthermore, the temperature gradient at the bottom boundary is fixed
to have a constant heat flux into the domain, and the latitudinal
boundaries are impermeable for heat fluxes.
On the upper radial boundary we either apply a black body condition,
\begin{equation}
\sigma T^4  = -K\frac{\partial T}{\partial r} - \chiS \rho T \frac{\partial s}{\partial r},
\label{eq:bbc}
\end{equation}
or a constant temperature
\begin{equation}
T  = T_{\rm ref}(r=\Rc).
\label{eq:ct}
\end{equation}
In the former case the heat is transported out of the domain via an
enhanced SGS flux; see \cite{KMCWB13} and in the latter case via a
cooling flux; see \cite{WKMB13}.

We apply roughly 10--50 times larger values of viscosity and
magnetic diffusivity in the coronal envelope compared to the convection
zone to avoid high velocity amplitudes and strong shear flows
resulting in small-scale magnetic field enhancements.
For the transition region we use a hyperbolic tangent function with a radial
width of $0.01\,R$ at $r=1.06\,R$, which is the largest value feasible.
In the case of Run~A5, the viscosity in the coronal envelope is
just three times higher than in the convection zone, but we apply an
additional shock
viscosity and shock diffusivity; see \cite{HBM04} and \cite{GSFSM13}
for details regarding their implementation.
The heat conductivity $K$ in the coronal envelope is chosen such that
the heat diffusivity $\chi=K/\rho c_{\rm P}$ is constant.

We characterize the runs by the values of the input
parameters: Prandtl number $\Pr=\nu/\chi^{m}$, sub-grid scale Prandtl number
$\PrS=\nu/\chiSm$, magnetic Prandtl number $\Pm=\nu/\eta$, Taylor
number $\Ta=(2\Omega_0 (R-r_0)^2/\nu)^2$, and Rayleigh number,
\begin{eqnarray}
\Ra\!=\!\frac{GM(R-r_0)^4}{\nu\,\chiSm\,R^2} \left(-\frac{1}{c_{\rm
    P}}\frac{\dd s}{\dd r} \right)_{r=0.85\, R},
\label{equ:Ra}
\end{eqnarray}
which is obtained from a hydrostatic one-dimensional model for the same initial
setup with $\chi^{m}=\chi(r=0.85\,R)$ and $\chiSm=\chiS(r=0.85\,R)$.
Furthermore we define the fluid and magnetic Reynolds numbers,
$\Rey=\urms/\nu\kf$ and $\Rm=\urms/\eta\kf$, respectively, the
Coriolis number $\Co=2\Omega_0/\urms\kf$ and the P\'eclet number
$\Pe=\urms/\chiSm\kf$,
where $\kf=2\pi/(R-r_0)\approx21/R$ is used as a reference wavenumber
and where $\urms$ is the typical turbulent velocity in the convection zone
defined as
\begin{equation}
\urms=\sqrt{3/2\brac{u_r^2+u_{\theta}^2}_{\theta\phi r\leq R}}\ ,
\end{equation}
which corrects for the removal of the differential rotation-dominated
$u_{\phi}$.
The slow mean meridional flows, however, are not removed.
Azimuthal averages combined with time averages in the saturated stage
are referred to as mean and indicated with an overbar, e.g., $\meanBB$,
while other averages are indicated as $\brac{.}$ with the spatial directions as
indices.
The index $0$ refers to the value at the bottom of the domain,
that is $\rho_0$ and $T_0$.
We also use the meridional distribution of turbulent velocities
\begin{equation}
\urmsp(r,\theta)=({\overline{\uu^{\prime\,2}}})^{1/2},
\end{equation}
where the fluctuating velocity is defined via
$\uu^\prime=\uu-\meanuu$.
Thus, meridional mean flows are here removed.
The mesh is equidistant in all directions. The grid resolutions are
given in \Tab{runs}.

We express our results in physical units following
\cite{KMCWB13,KMB14} and \cite{WKKB14} by choosing a normalized
rotation rate $\Ot$ =$\Omega_0/\Omega_{\odot}$, where
$\Omega_{\odot}=2.7\times10^{-6}\,$s$^{-1}$ is the solar rotation rate.
The simulations were performed with the {\sc Pencil
  Code}\footnote{http://github.com/pencil-code/}, which uses a
high-order finite difference method for solving the compressible
equations of magnetohydrodynamics.

\begin{figure}[t!]
\begin{center}
\includegraphics[width=0.9\columnwidth]{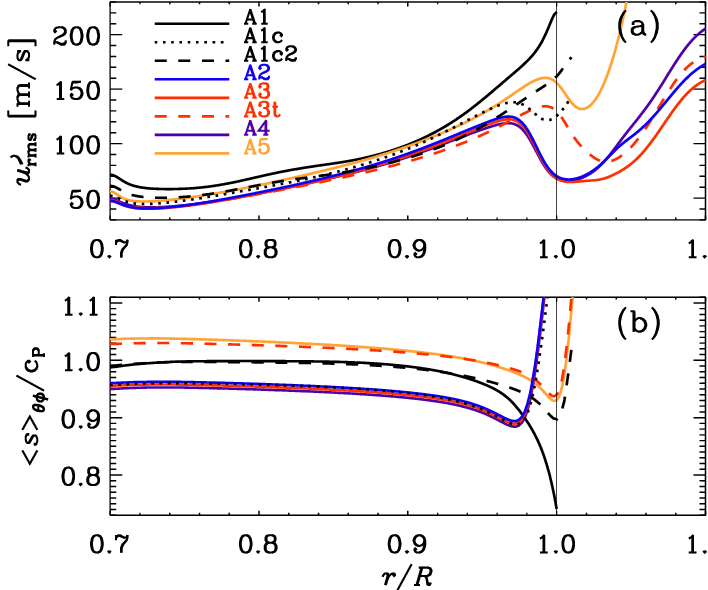}
\includegraphics[width=0.9\columnwidth]{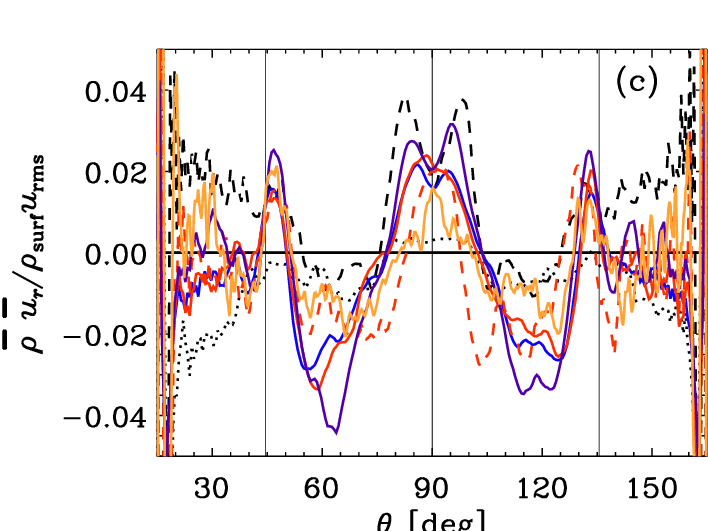}
\end{center}\caption[]{
Radial profiles of azimuthally and latitudinally averaged turbulent rms
velocity $\urmsp$ (a) and $\brac{s}_{\theta\phi}$ (b) in m/s or normalized by
$c_P$ respectively for Set~A.
Mean radial mass flux $\meanrho\,\mean{u_r}/\rho_{\rm surf}\urms$ (c)
through the surface ($r=R$) for Set~A.
The horizontal line indicates the zero value and
the three vertical thin lines indicate the equator
($\theta=90^\circ$) and the intersection with the inner tangent cylinder
($\theta-90^\circ\approx\pm45^\circ$ latitude).
}
\label{strat2_A}
\end{figure}

\section{Results}

\begin{table*}[t!]\caption{
Summary of runs.
}\vspace{12pt}\centerline{\begin{tabular}{lccccccccccccc}
Run & Resolution  & $\Ot$ & $\Rc/R$& $\Pr$ & $\PrS$ & $\Pm$ & $\Ta$ &$\Ra$ &$\rho_0/\rho_{\rm
  surf}$& $\rho_0/\rho_{\rm top}$ &$\Co$ & $\Rey$ & $\Pe$ \\
\hline
\hline
A1 &   $180\times256\times128$ & 5 & 1.0   & 73&2& 1& 1.25$\,\cdot 10^{8}$ & 4.0 $\,\cdot
10^{7}$& 31 & 31 & 8.1 & 35 & 69\\ 
A1c &  $128\times256\times128$ & 5 & 1.01 & 78& 2& 1& 1.25$\,\cdot 10^{8}$ & 3.9$\,\cdot
10^{7}$& 51 & 80 & 9.4 & 30 & 60\\ 
A1c2 & $128\times256\times128$ & 5 & 1.01& 68& 2& 1& 1.25$\,\cdot 10^{8}$ & 2.9$\,\cdot
10^{7}$& 22 & 31 & 9.4 & 30 & 60 \\ 
A1pc & $180\times256\times128$ & 5 & 1.0  & 73& 2& 1& 1.25$\,\cdot 10^{8}$ & 4.0$\,\cdot
10^{7}$& 35 & 35 & 8.3 & 34 & 68 \\ 
\hline
A2 &   $290\times256\times128$ & 5 & 1.2 & 79& 2& 1& 1.25$\,\cdot 10^{8}$ & 4.0$\,\cdot
10^{7}$& 59 & 7970 & 10.4 & 27 & 54 \\ 
A3 &   $400\times256\times128$ & 5 & 1.4 & 79& 2& 1& 1.25$\,\cdot 10^{8}$ & 3.9$\,\cdot
10^{7}$& 61 & 10510 & 10.1 & 27 & 53\\  
A3t &  $400\times256\times128$ & 5 & 1.4 & 67& 2& 1& 1.25$\,\cdot 10^{8}$ & 3.0$\,\cdot
10^{7}$& 23 & 5380 & 10.2 & 28 & 55\\  
A4 &   $520\times256\times128$ & 5 & 1.6 & 79& 2& 1& 1.25$\,\cdot 10^{8}$ & 3.9$\,\cdot
10^{7}$& 59 & 12740 & 10.8 & 26 & 52\\  
\hline
A5 &   $600\times512\times256$ & 5 & 1.4 &35&0.5&0.5& 5.00$\,\cdot
10^{8}$ &1.8$\,\cdot 10^{7}$ & 22 & 4937 & 8.7 &65 & 32\\  
\hline
\hline
B1 &   $180\times256\times128$ & 3 & 1.0 & 73& 2& 1& 0.45$\,\cdot 10^{8}$ & 4.0$\,\cdot
10^{7}$& 36 & 36 & 4.3 & 39 & 79 \\
B1c &  $128\times256\times128$ & 3 & 1.01 & 78& 2& 1& 0.45$\,\cdot 10^{8}$ & 3.9$\,\cdot
10^{7}$& 52 & 83 & 5.6 & 33 & 67 \\
B3 &   $400\times256\times128$ & 3 & 1.4 & 78& 2& 1& 0.45$\,\cdot 10^{8}$ & 3.9$\,\cdot
10^{7}$& 58 & 8803 & 5.5 & 31 & 61\\
B3t &  $400\times256\times128$ & 3 & 1.4 & 79& 2& 1& 0.45$\,\cdot 10^{8}$ & 3.0$\,\cdot
10^{7}$& 24 & 5380 & 5.4 & 32 & 64 \\
\hline
\label{runs}\end{tabular}}\tablefoot{
The second to ninth columns show quantities that are input parameters
to the models, whereas the quantities in the last five columns are
results of the simulations computed from the saturated state.
All quantities are volume averaged over the convection zone $r\leq R$,
unless explicitly stated otherwise.
$\tilde{\Omega}=\Omega_0/\Omega_{\odot}$ is the normalized rotation
rate and $\Rc/R$ is the outer radius of the domain.
Here, $\rho_0$, $\rho_{\rm surf}$, and $\rho_{\rm top}$ are the latitudinal
and azimuthally averaged density at the bottom ($r=0.7\,R$), the surface
($r=R$), and the top ($r=\Rc$) of the domain.
}
\end{table*}

In this work we compare and analyze 13 runs divided into two sets based
on the rotation rate. 
In Set~A the normalized rotation rate is $\Ot=5$ whereas the runs in Set~B have
a rotation rate of $\Ot=3$.
For both sets we investigate the effects of a cooling layer and the
blackbody boundary condition as well as the size of the coronal
envelope.
A summary of the runs can be found in \Tab{runs} and the
stratifications of temperature, density and entropy of Runs~A are
shown in \Figs{strat1_A}{strat2_A}.
Run~A1 is nearly the same as Run~B4m of \cite{KMB12},
Run~C1 of \cite{KMCWB13}, Run~I of \cite{WKKB14}, and the run of
\cite{WRKKB16}, where a blackbody boundary condition is used.
However, we choose a slightly higher stratification and a slightly
lower value of $\PrS$, namely 2 instead of 2.5; therefore Run~A1 is the same as
Run~D3 of \cite{KKOWB16}.
In Runs~A1c and A1c2, the blackbody boundary condition is replaced by
a shallow ($R\le r\le\Rc=1.01\,R$) cooling layer, where in the former case
the temperature minimum is below the surface ($r=R$) and in the latter
above the surface.
These two runs have been recently used in \cite{WKKB14} as Runs~III and
IV, respectively.
The only difference between Runs~A1pc and A1 is the use of a perfect
conductor condition instead of a radial field condition for the
magnetic field at the top boundary.

The other runs of Set~A have a coronal envelope with different
outer radii $\Rc$.
Runs~A2, A3, and A4 have the same cooling function as Run~A1c, where the
temperature reaches a minimum below the surface.
The temperature increases to a constant coronal value, which is more
than twice the value at the bottom of the convection
zone; see \Fig{strat1_A}(a).
This results in a positive entropy gradient above $r=0.97\,R$, where the
convection ceases and $\urmsp$ drops by a factor of two; see
\Fig{strat2_A}(a,b).
In Run~A3t, the same cooling function is applied as in Run~A1c2,
leading to a temperature minimum above the surface at $r\approx1.01\ R$.
This causes the entropy gradient to become positive at the surface
($r=R$) and an increase of $\urmsp$ all the way to the surface; see
\Fig{strat2_A}(a,b).
Already here, we can state that the use of cooling profiles in Runs~A1c and A3t
reproduce most properties of the density, temperature and entropy
stratification as the blackbody boundary conditions in Run~A1.
In Run~A5, we lower $\PrS$ and $\Pm$ to $0.5$ and therefore
increase the fluid Reynolds number.
As the heat flux at the boundary is the same as in the other runs,
$\urmsp$ increases only slightly, see \Fig{strat2_A}(a), leading to a
reduced Coriolis number; otherwise Run~A5 is similar to Run~A3t.
However, the use of a lower value of viscosity in combination with a
shock viscosity allows higher velocities in the coronal envelope as
shown in \Fig{strat2_A}(a).
The runs of Set~B are essentially the same as the corresponding runs in
Set~A with a lower rotation rate. 
The radial temperature, density, entropy and velocity
profiles behave similarly as in Set~A and are therefore not
shown here.
In the following we investigate the influence of the coronal layer on
mass flux and temperature distribution (\Sec{sec:flux}) as well as on
differential rotation and meridional circulation (\Sec{sec:diff}).
Furthermore we discuss Reynolds stresses and the $\Lambda$ effect
(\Sec{sec:rey}) and their contribution to the thermal wind balance
(\Sec{sec:therm}).
Then we investigate the influence of the magnetic top boundary on the
field structure near the surface (\Sec{sec:radfield}) and the magnetic
field evolution (\Sec{sec:dynamo}).

\subsection{Mass flux and temperature distribution at the boundary}
\label{sec:flux}

\begin{figure}[t!]
\begin{center}
\includegraphics[width=0.9\columnwidth]{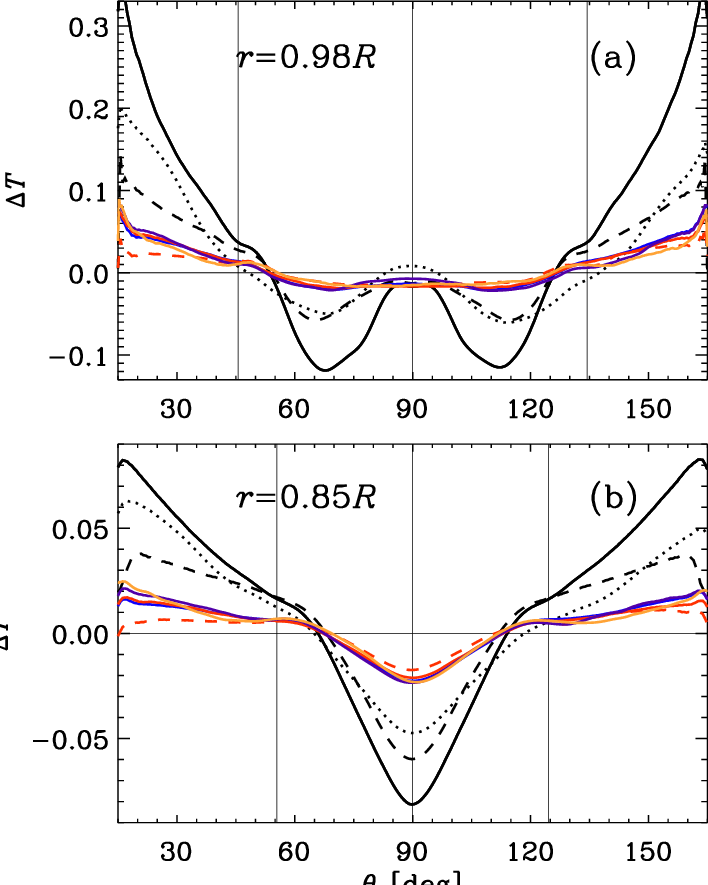}
\end{center}\caption[]{
Latitudinal dependence of 
$\Delta T=(\mean{T}-\brac{\mean{T}}_{\theta})/\brac{\mean{T}}_{\theta}$
at radius $r=0.98\,R$ (a) and at radius $r=0.85\,R$ (b) for
Set~A. Linestyles as in \Fig{strat1_A}.
The thin black lines indicate the zero value, the equator
($\theta=\pi/2$) and the location of the inner tangent cylinder.
}
\label{Temp_A}
\end{figure}

We begin by inspecting the influence of the top boundary on the mass
flux and the temperature distribution; see \Fig{strat2_A}(c).
Except for Run~A1, the mass flux at $r=R$ is nonvanishing, showing a strong
latitudinal dependency.
Near the equator it is positive (outflow) and at
latitudes around $\pm$30$^{\circ}$ negative (inflow) suggesting a circulation in
the coronal envelope.
At mid-latitudes the mass flux becomes positive again, but is fluctuating
around zero toward higher latitudes.
The flow structure is strongly influenced by the rotation, as seen from
the alignment with the inner tangent cylinder; see \Fig{strat2_A}(c).
Although the mass flux in the runs with an extension above the
surface has non-zero values, they are small in comparison to
$\meanrho\urms$ at the surface.
Furthermore, there is no qualitative difference between runs with a
cooling layer $\Rc=1.01\,R$ (Runs~A1c and A1c2) and a coronal
envelope $\Rc\ge1.2\,R$ (Runs~A2, A3, A3t, and A4). In fact, Run~A1c2
has the highest mass flux through the surface; see \Fig{strat2_A}(c).
Furthermore, Run~A5 is similar to the other runs showing more
fluctuations as a function of latitude but with a similar magnitude of
variations as in the other runs.
This shows that the influence of a coronal envelope via a
radial mass flux is small and can be neglected.
Furthermore, we find no indication that the viscosity profile has a major
influence on the mass flux.
 
As a second step we investigate the latitudinal temperature
variation, $\Delta
T=(\mean{T}-\brac{\mean{T}}_{\theta})/\brac{\mean{T}}_{\theta}$, at two radii; see
\Fig{Temp_A}.
In general the surface perturbations are strongest near the poles,
decrease toward a minimum at mid-latitudes, and increase again below
$\pm20\degr$ latitude. However, in the middle of the convection zone 
the temperature minimum is at the equator.
This is a clear indication of a strong rotational influence on
the temperature distribution.
Run~A1 shows the largest relative temperature perturbations; over
0.3 near the poles and up to $-0.1$ at $\pm20\degr$ latitude.
Here, the radiative boundary condition lets the temperature at
the surface evolve more freely, which leads to this strong variation.

Runs~A1c and A1c2 have a similar distribution, but with around three times
smaller values.
There the cooling layers cool the temperature to a
certain latitudinally independent value with a relaxation time
equal to the turnover time.
This leads to a reduction of the temperature perturbations near the
surface.
The temperature difference $\Delta T$ is significantly reduced for all
runs with a coronal envelope.
However, the temperature is still higher near the poles and lower near
the equator than the latitudinal average.
In these runs the coronal envelope with its mass and heat
capacity serves as a buffer in smoothing the temperature at the
surface.
The influence of the cooling layer and the coronal envelope seems to
penetrate also deeper down and influences the temperature variations in
the middle of convection zone.
The difference in the temperature profiles caused by the coronal
envelope can also influence differential rotation; see
\Sec{sec:diff}.

In the higher $\Rey$ and thus more turbulent Run~A5, the
temperature variation is similar to that of the other runs with a coronal
envelope.
The runs with slower rotation (Set~B, not shown here) show a similar behavior, but
the latitudinal temperature perturbations are weaker than
in the more rapidly rotating runs, due to the reduction of the rotational
influence.

On the solar surface, systematic latitudinal temperature variations have
not been observed.
However, variations in the range of a few kelvin, so less than 0.1\% of
the surface temperature, are below the measurable range.
This value is exceeded in all of our simulations.
The temperature difference between the poles and equator can influence
the differential rotation in the Sun \citep{R89} and in simulations
\citep[e.g.,][]{MBT06,WKMB13}.

\subsection{Differential rotation and meridional circulation}
\label{sec:diff}

\begin{figure}[t!]
\begin{center}
\includegraphics[width=0.242\textwidth]{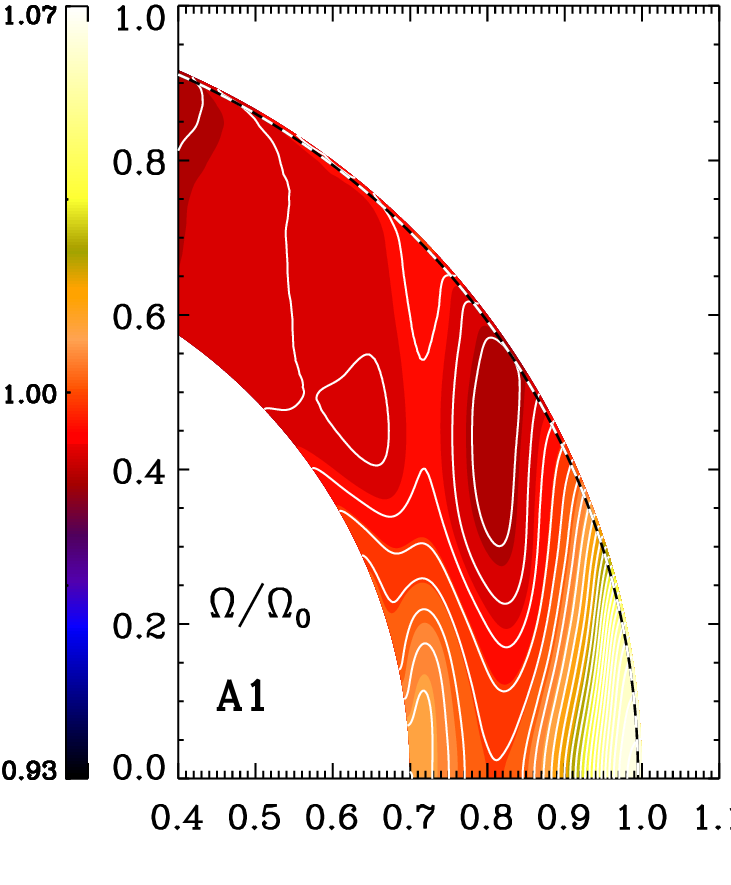}
\includegraphics[width=0.242\textwidth]{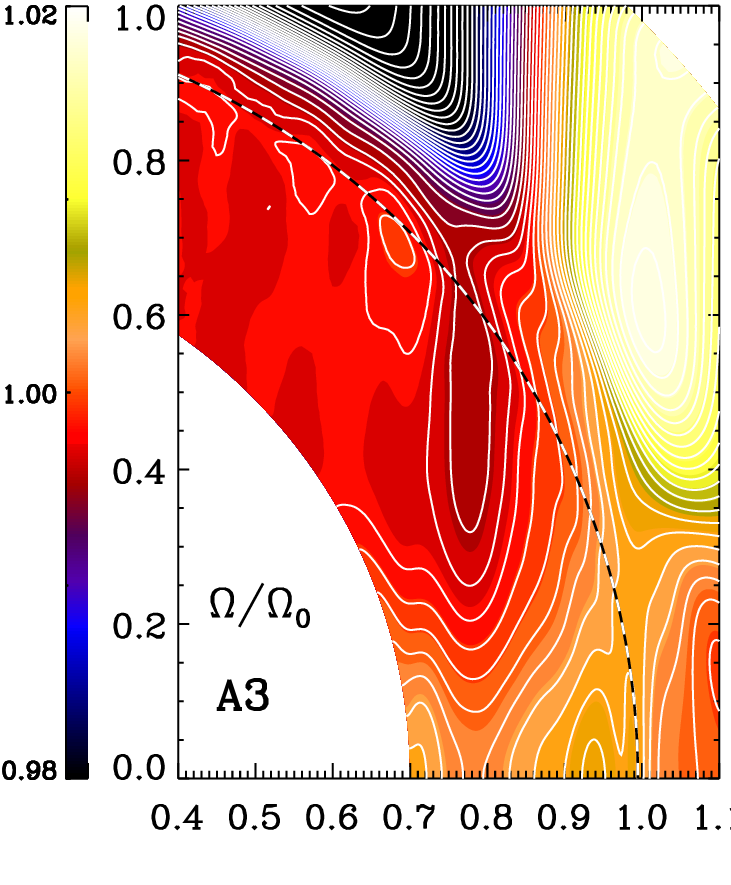}
\includegraphics[width=0.242\textwidth]{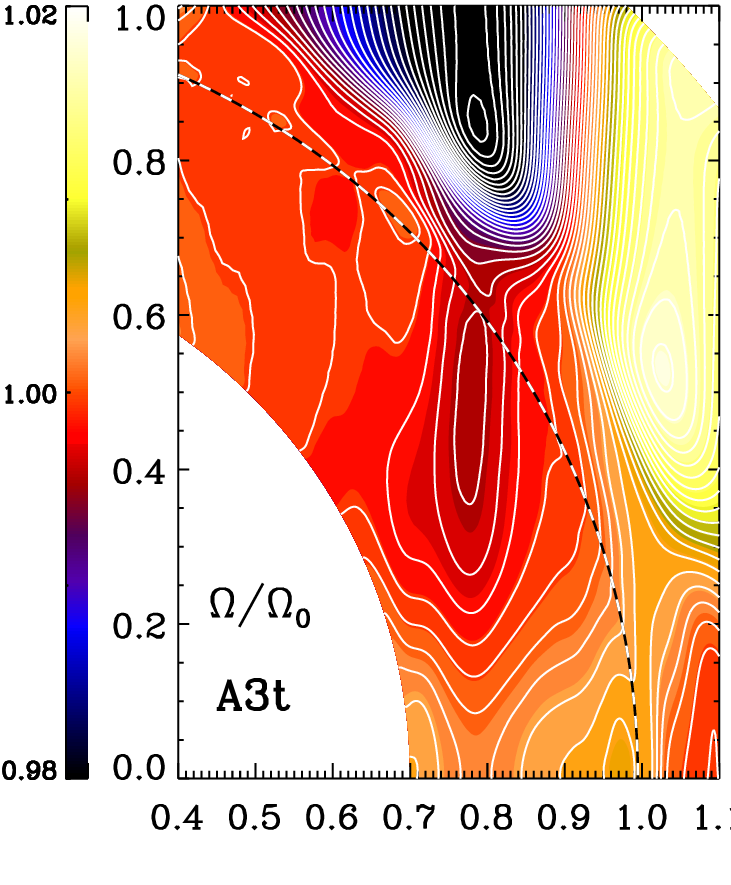}
\includegraphics[width=0.242\textwidth]{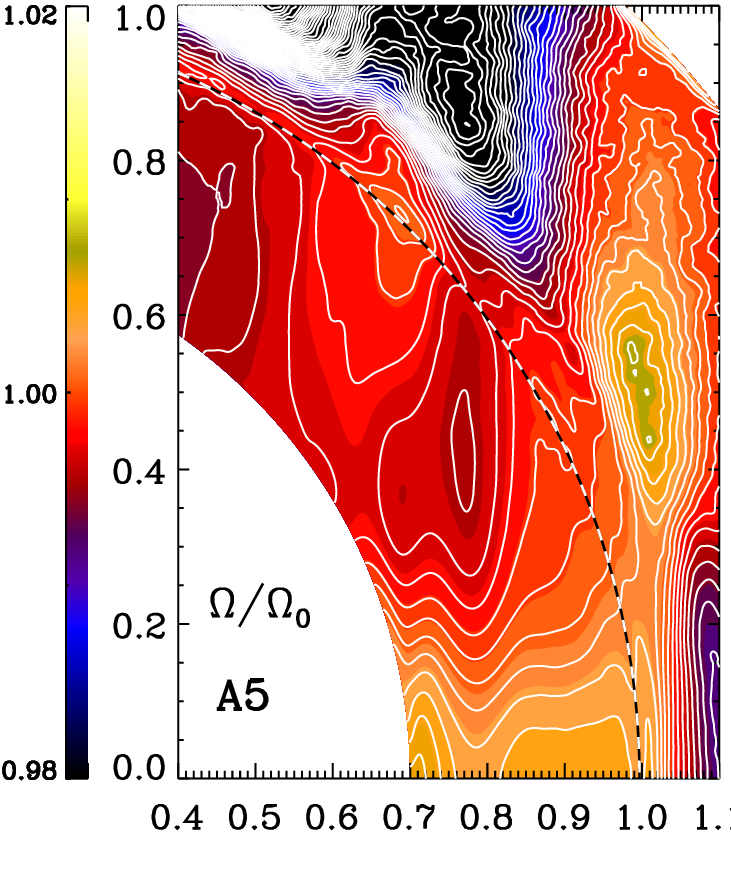}
\includegraphics[width=0.242\textwidth]{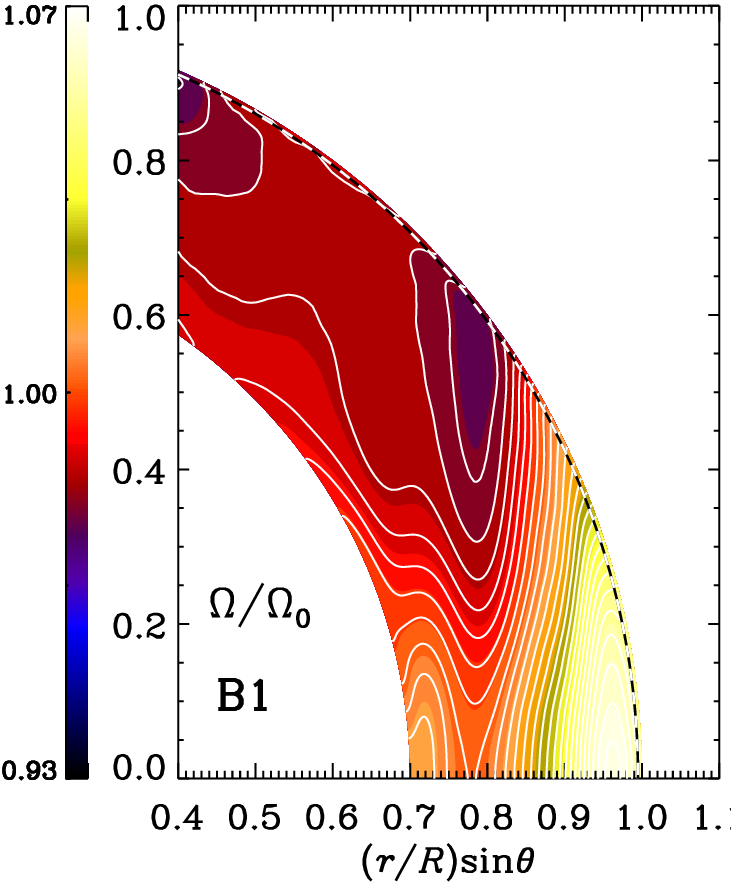}
\includegraphics[width=0.242\textwidth]{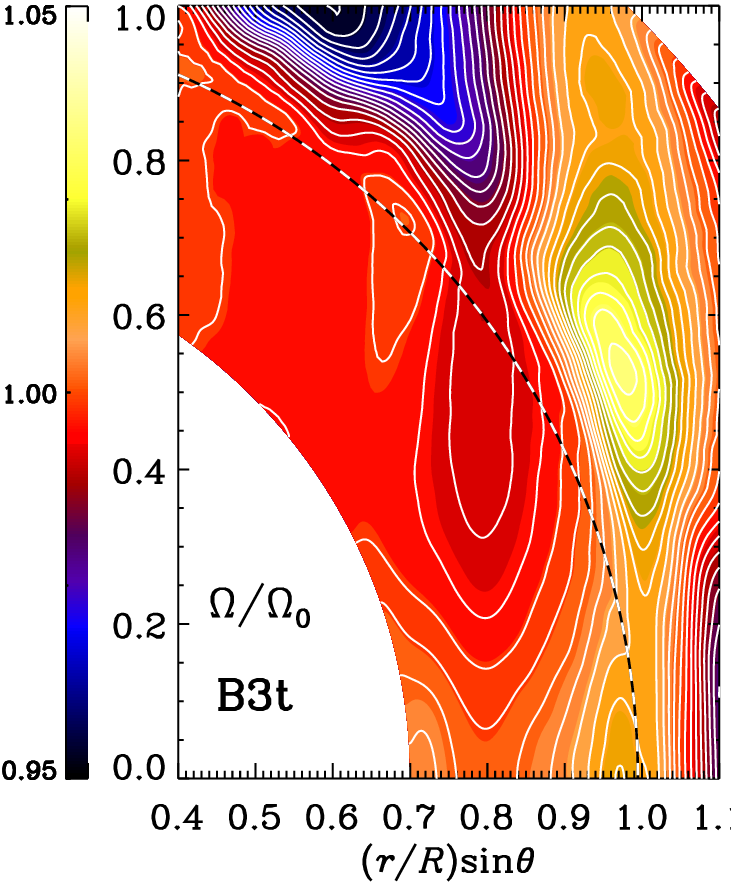}
\end{center}\caption[]{
Angular velocity $\Omega(r,\theta)/\Omega_0$ for Runs~A1 and A3 ({\it top
row}), Runs~A3t and A5 ({\it middle row}) and Runs~B1 and B3t ({\it bottom row}).
Here, $\Omega=\Omega_0+\mean{u_\phi}/r\sin\theta$ is the local
rotation rate. The dashed lines indicate the surface ($r=R$).
}
\label{diff}
\end{figure}

\begin{figure}[t!]
\begin{center}
\includegraphics[width=\columnwidth]{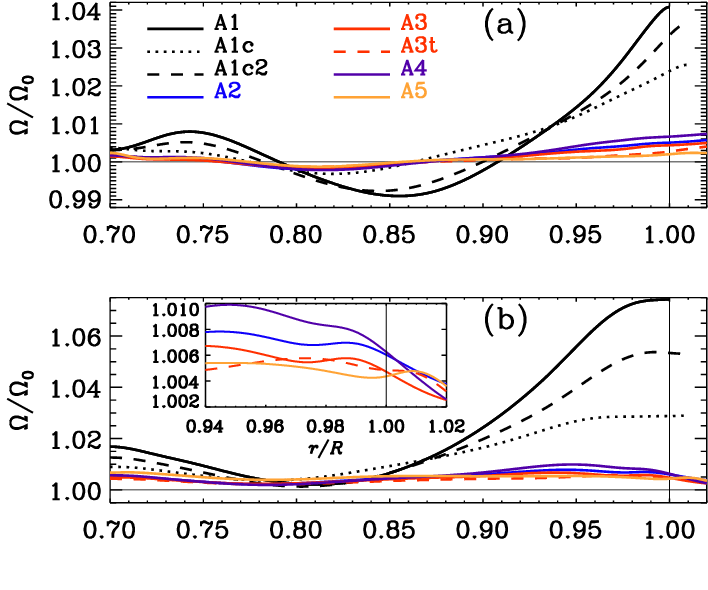}
\includegraphics[width=\columnwidth]{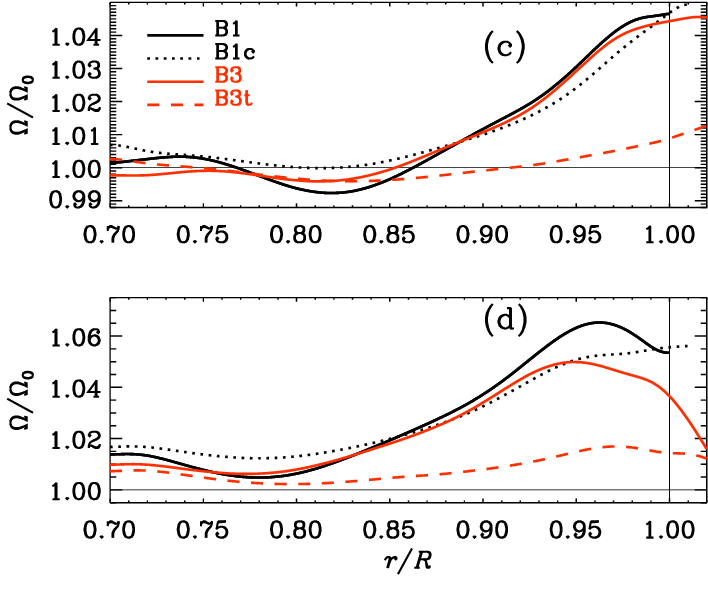}
\end{center}\caption[]{
Differential rotation $\Omega(r,\theta)/\Omega_0$ as a function of 
radius for all runs of
Set~A at $\theta=75^\circ$ (15$^\circ$ latitude) (a), and at the
equator
$\theta=\pi/2$ (b), and Set~B at mid-latitudes $\theta=75^\circ$ (c),
and at the equator $\theta=\pi/2$ (d). 
The inlay in (b) shows the angular velocity near the surface for Runs~A2, A3,
A3t, A4, and A5.
}
\label{diff_cut}
\end{figure}

\begin{figure}[t!]
\begin{center}
\includegraphics[width=0.242\textwidth]{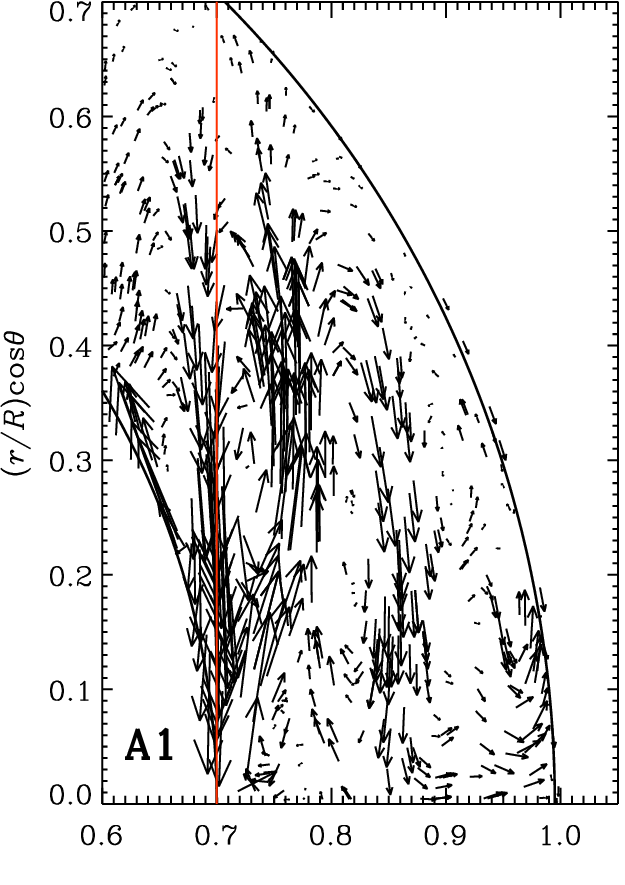}
\includegraphics[width=0.242\textwidth]{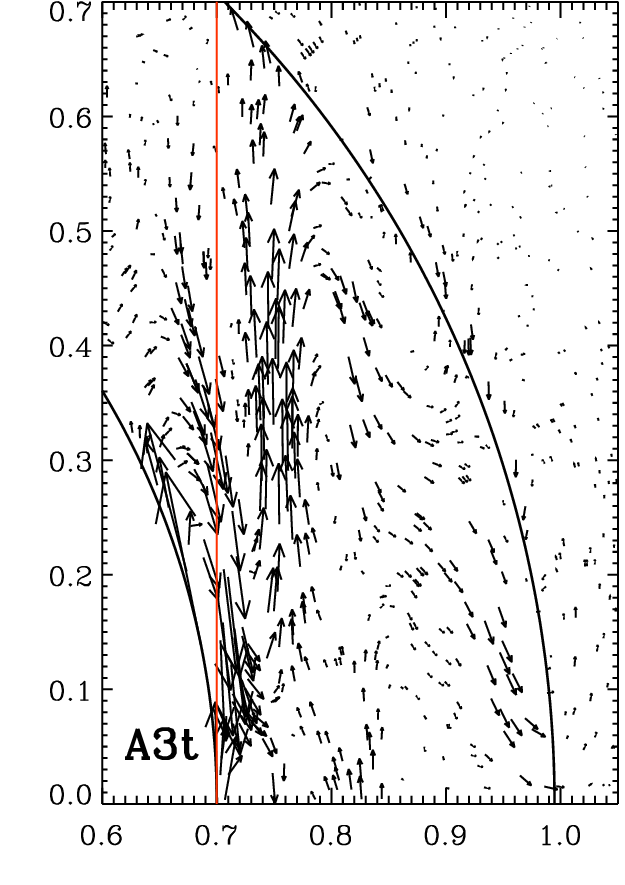}
\end{center}\caption[]{
Meridional circulation in terms of the mass flux
$\meanrho\,\mean{\uu_m}$ for Runs~A1 and A3t.
The dashed lines indicate the surface ($r=R$) and the red solid line
the inner tangent cylinder.
}
\label{meri}
\end{figure}

As shown in \cite{WKMB13}, simulations with a coronal envelope
are more capable of reproducing a spoke-like differential rotation
profile than runs without a coronal envelope using similar
parameters \citep{KMCWB13}.
In this work, the parameters of the runs are nearly identical,
so we can isolate the influence of the coronal envelope.
In \Fig{diff} we plot the angular velocity,
$\Omega=\Omega_0+\mean{u_\phi}/r\sin\theta$, in the meridional plane
and in \Fig{diff_cut} as a function of radius at two latitudes.
In all of the runs the equator rotates faster than the poles, similarly
to the other simulations with similar Coriolis numbers
\citep{BBBMT10,KMCWB13,WKMB13,ABMT15}.
Furthermore, all runs possess a local minimum in the
angular velocity $\Omega$ at mid-latitudes, which has been shown to 
facilitate equatorward migration \citep{WKKB14,KKOWB16}.
In runs without coronal envelopes the rotation profile has a similar
structure; the contours of constant angular velocity show a strong alignment
with the rotation axis, following the Taylor-Proudman theorem.
Run~A1 possesses the strongest differential rotation.
In particular at the equator the surface rotates faster than in the
other runs.
Runs~A1c and A1c2 show a similar radial dependency, but differential rotation is
weaker than in Run~A1; see \Fig{diff_cut}(a) and (b).

In the runs with a coronal envelope the Taylor-Proudman balance is
broken and the rotation profile has a more spoke-like shape; see
\Fig{diff}.
The differential rotation and the local minimum at mid-latitudes in
these runs is much weaker. The minimum also occurs at a greater depth
than in the runs without a corona.
The dependence on the size of the coronal extent on differential
rotation is weak -- similarly to what is seen for the latitudinal
temperature distribution; see \Fig{Temp_A}.
The cooling profile has a minor influence on the rotation profiles
such that in Run~A3t the contours of constant rotation are slightly
more radial than in Run~A3
In Run~A5, the differential rotation is even weaker with
more radial contours of rotation.

The overall rotation profiles of Set~B are similar to those of Set~A
and also show a local minimum of $\Omega$ at mid-latitudes although it
is weaker and at a greater depth; see \Fig{diff}.
Also here, the presence of a coronal envelope leads to more radial
contours of rotation.
However, the amount of differential rotation is reduced only for
Run~B3t, but not for Run~B3.

If we compare the differential rotation profiles with those of
\cite{WKMB13}, we see that the contours of constant rotation are more
cylindrical here.
In particular, moderate rotation runs had more
spoke-like differential rotation than rapidly rotating ones which is
not seen in the present work.
Therefore, we relate the more spoke-like rotation profiles in \cite{WKMB13} to a
lower density stratification as the other parameters are similar.

For most of the runs (A2, A3, A3t, A4, B1, B3, and B3t) the maximum of rotation
at the equator is actually below the surface, indicating a
near-surface shear layer with negative radial shear.
The logarithmic gradient of rotation, $\dd \ln \Omega/\dd \ln r$, is around
$-0.2$ near the surface for Run~A4, and $-0.15$ for Runs~A2, A3, and A3t.
This is much weaker than the value for the Sun, which is $\dd \ln
\Omega/\dd \ln r\approx-1$ for all latitudes \citep{BSG14,BSG16}.
In all runs of Set~B, except for B1c, the near-surface shear region is more
extended than the ones in Set~A, and the gradient is stronger; $\dd
\ln \Omega/\dd \ln r$ reaches values of $-0.8$ for Run~B3 and $-0.5$
for Run~B3t.
It is expected that for runs with lower rotation rate, a near-surface
shear layer is stronger due to the weaker influence of the Coriolis
force near the surface; see \Secs{sec:rey}{sec:therm}.
In agreement with $\Lambda$ effect theory \citep{R80,R89}, the
double-logarithmic gradient should only be close to $-1$ very near the
surface where the local Coriolis number is small
\citep{KR05,K13,RKT14,K16}.
While this is true for the Sun, it is not the case in
our simulations owing to limited stratification.

To investigate the influence of the cooling profile on the temperature
variation and differential rotation, we have performed two additional
runs, in which we either increased or decreased the cooling
luminosity compared to Run~A1c2 (not shown).
A decrease of the cooling luminosity by a factor of two leads to a shift
of the temperature minimum at higher radii.
The mean temperature increases slightly, resulting in a higher density
at the surface and a decrease
of the density stratification in the convection zone.
An increase of the cooling luminosity has the opposite effect, leading to
a temperature minimum
at a greater depth, and a lower temperature and density in the
convection zone.
Weaker cooling causes stronger differential rotation, especially
at higher latitudes, while stronger cooling does not show a
significant effect.
The gradient $\dd \ln \Omega/\dd \ln r$ at the equator becomes more
negative with a weaker cooling.

Differential rotation is also generated in the coronal envelopes.
Below $r=1.01\,R$, the differential rotation follows the rotational
behavior of the convection zone.
Above $r=1.01\,R$ the plasma rotates nearly uniformly near the equator
with a rotational speed close the $\Omega_0$.
The mid-latitudes rotate faster than the equator and at high latitudes
the coronal envelopes decrease to slower rotation.
All runs exhibit strong cylindrical and radial shear.
This is consistent with \cite{WKMB13}, where runs with lower
stratification showed a similar behavior.
The change of behavior from below to above $r=1.01\,R$ is caused by
the change in temperature and density stratifications and not by the
viscosity profiles.
This can be seen by comparing Runs~A3t and A5, where the same
temperature and density profiles are used, but a much lower viscosity
is applied in Run~A5 than in Run~A3t.

In \Fig{meri} we plot the meridional circulation in terms of the mass
flux $\meanrho\,\mean{\uu_m}$ in the meridional plane, where
$\mean{\uu_m}=(\mean{u_r},\mean{u_\theta},0)$ is the meridional flow.
The meridional circulation has a multi-cellular structure in all runs.
Near the equator at the surface the flow is poleward, but it can become
equatorward at high latitudes; see \Fig{meri}.
The strongest contribution to the mass flux carried by the meridional
circulation occurs within the bulk of the convection zone.
There the flow is aligned with the rotation axis and streaming
toward the equator along the inner tangent cylinder and toward higher latitudes
further away from the rotation axis.
These mass flows seem to stream toward the local minima of $\Omega$ at
mid-latitudes.
From there, most of the runs develop a flow toward the equator following
the $\theta$ direction.
The stronger meridional flows in Run~A1 are due to the higher density, see
\Fig{strat1_A}(b), while the actual flow is quite similar in all runs of Set~A,
see \Fig{Rey_A}(h).
The runs of Set~B generate stronger meridional circulation, similarly
to what was found in \cite{WKMB13}.
At these rotation rates, slower rotation leads to an increase of meridional
circulation as found in mean-field models \citep{Ko70,R89} and numerical simulations
\citep[e.g.,][]{BBBMT08,ABBMT12}.
In general the meridional flow pattern does not change due to the
influence of the coronal envelope.

\subsection{Reynolds stresses and $\Lambda$ effect}
\label{sec:rey}
\begin{figure*}[t!]
\begin{center}
\includegraphics[width=18cm]{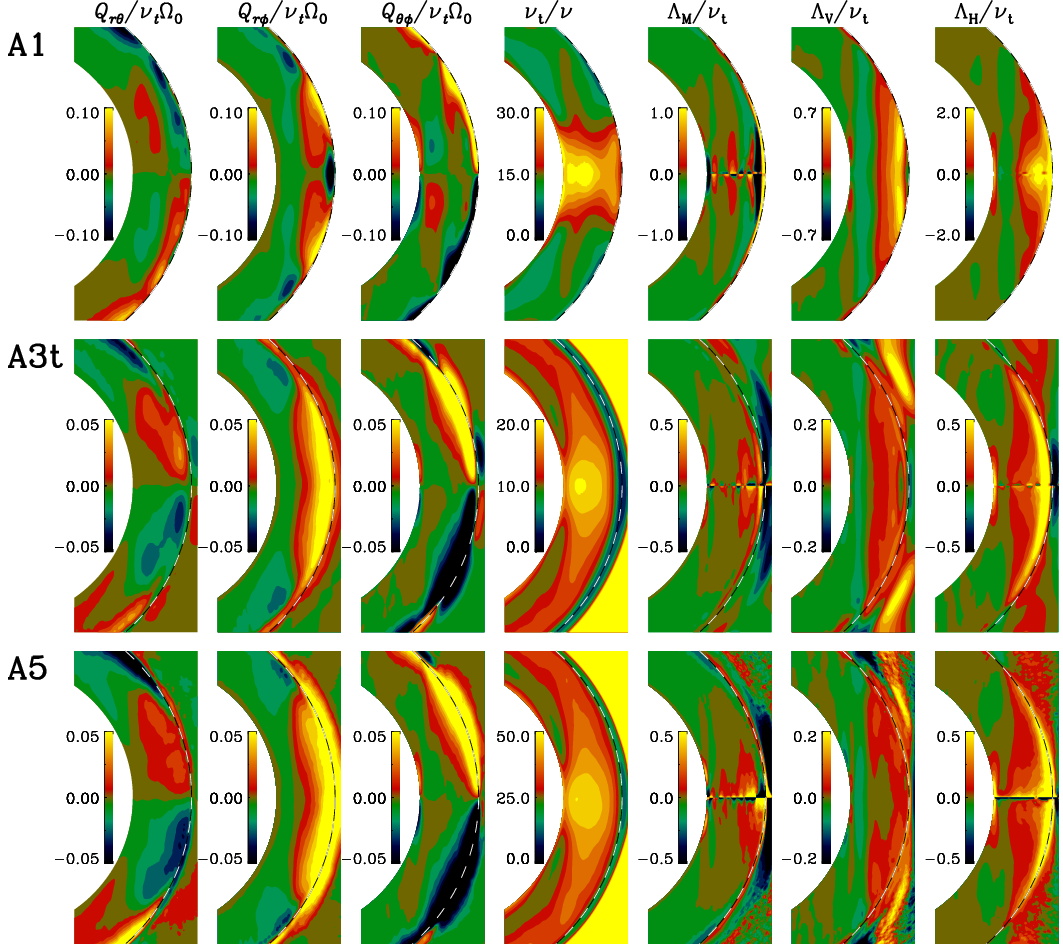}
\end{center}\caption[]{
From left to right: Off-diagonal components of the Reynolds stress
$Q_{r\theta}$,
$Q_{r\phi}$ and $Q_{\theta\phi}$ normalized by $\nu_t\Omega_0$ (first
three columns), the turbulent viscosity in terms of molecular
viscosity $\nu_t/\nu$ (fourth column) and the three components of the
$\Lambda$ effect $\LLM$, $\LLV$, and $\LLH$ (fifth to
seventh column) normalized by $\nu_t$ for Runs~A1 (top row), A3t
(middle) and A5 (bottom).
}
\label{Rey_mer}
\end{figure*}

\begin{figure*}[t!]
\begin{center}
\includegraphics[width=\textwidth]{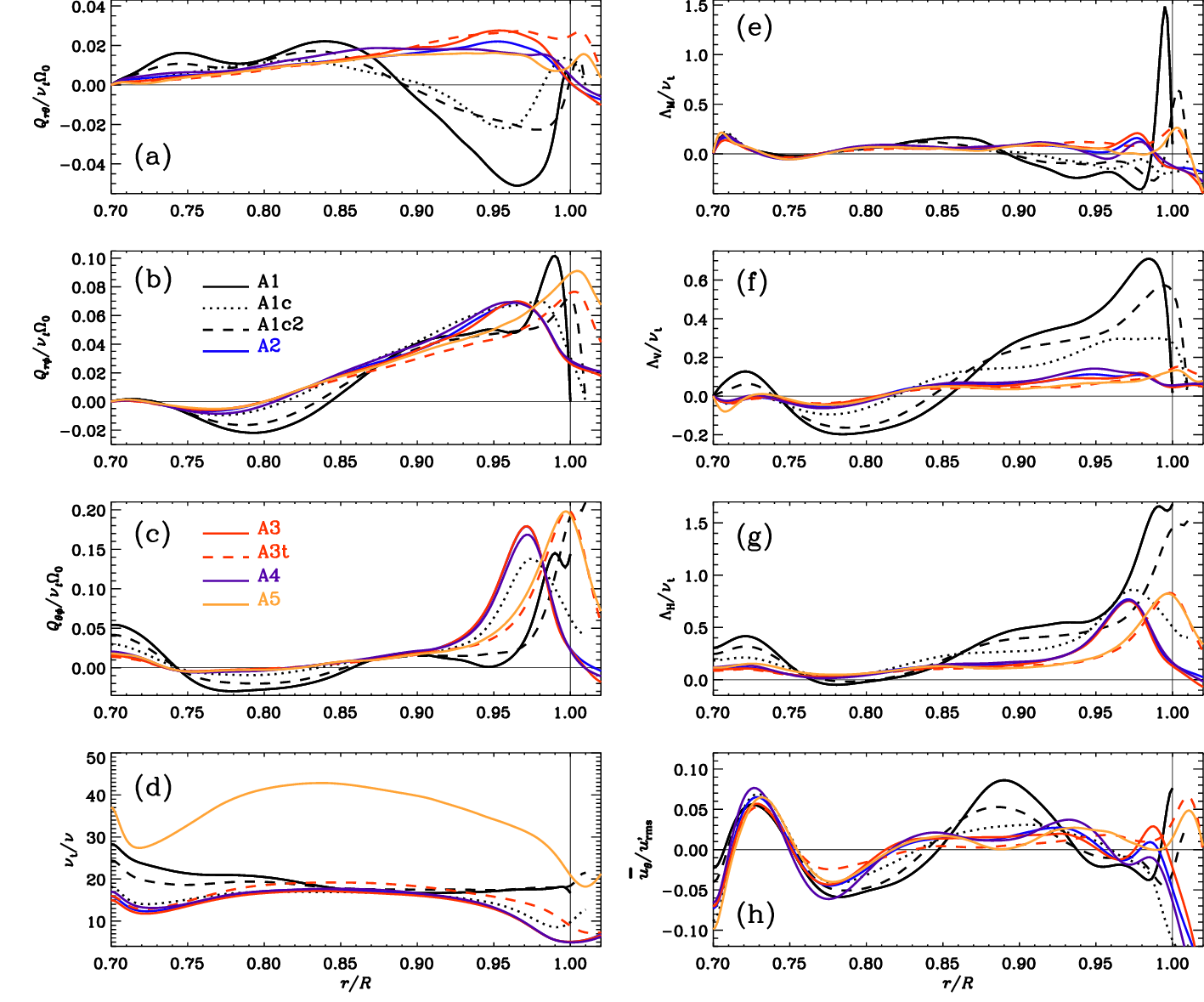}
\end{center}\caption[]{
Off-diagonal components of the Reynolds stress $Q_{r\theta}$ (a),
$Q_{r\phi}$ (b) and $Q_{\theta\phi}$ (c) normalized by
$\nu_t\Omega_0$, the turbulent viscosity in terms of molecular 
viscosity $\nu_t/\nu$ (d) and the three components of the
$\Lambda$ effect $\LLM$ (e), $\LLV$ (f) and $\LLH$ (g)
normalized by $\nu_t$ as well as the meridional flow
$\mean{u_\theta}/\urmsp$ (h) for Set~A in the northern hemisphere at
15$^\circ$ latitude.
The thin black lines indicate the zero value and the surface ($r=R$).
}
\label{Rey_A}
\end{figure*}

Differential rotation and meridional circulation in the Sun and other
stars is generated by the
interaction of turbulent convection and rotation \citep{R89}.
Reynolds stresses become anisotropic due to an angle between the
gravity and the axis of rotation.
The non-diffusive contribution of the Reynolds stress tensor $Q_{ij}$
can be expressed via the $\Lambda$ effect which produces equatorial
acceleration if the angular momentum transport is directed equatorward
\citep[see e.g.,][]{R89,KR95,KR05}.
It has been recently shown that there is strong evidence for the
$\Lambda$ effect operating in the Sun, causing the observed rapidly
rotating equator \citep{RKT14} and the latitude-independent surface
shear \citep{K16}.

We calculate the three off-diagonal components of the Reynolds stress
tensor $Q_{r\theta}=\overline{u_r^\prime u_\theta^\prime}$,
$Q_{r\phi}=\overline{u_r^\prime u_\phi^\prime}$, and
$Q_{\theta\phi}=\overline{u_\theta^\prime u_\phi^\prime}$.
Both $Q_{r\phi}$ and $Q_{\theta\phi}$ contribute to the angular momentum
balance and their non-diffusive parts are associated with the horizontal
and vertical $\Lambda$ effects, respectively.
Even though $Q_{r\theta}$ does not directly contribute to angular
momentum
transport, it has been argued to be important in generating a
near-surface shear
layer in global convection simulations \citep{HRY15}.
The Reynolds stresses can be written as \citep[e.g.,][]{R89,RKH13}
\begin{equation}
Q_{r\phi}=\LLV \sin{\theta}\,\Omega-\nut r \sin{\theta}\,
\frac{\partial\Omega}{\partial r},\label{laV}
\end{equation}
\begin{equation}
Q_{\theta\phi}=\LLH \cos{\theta}\,\Omega-\nut \sin{\theta}\,
\frac{\partial\Omega}{\partial \theta},
\label{laH}
\end{equation}
where $\LLV$ and $\LLH$ are the vertical and horizontal
components of the $\Lambda$ effect and $\nut$ is the turbulent
viscosity (assumed isotropic).
Following \cite{KMB14} we approximate $\nut$ as
\begin{equation}
\nut=\onethird \urmsp\alpha_{\rm MLT} H_p,
\end{equation}
where $\alpha_{\rm MLT}=5/3$ has been assumed for the mixing length parameter,
and $H_p=-[\partial \ln p(r,\theta)/\partial r]^{-1}$ is the pressure
scale height.

A meridional $\Lambda$ effect also exists \citep{PTBNS93,RBMD94,KKT04,KB08}
which is related to $Q_{r\theta}$ via
\begin{equation}
Q_{r\theta}=\LLM\sin{\theta}\cos\theta\, \Omega-\nut\left[{1\over
    r}{\partial \mean{u_r}\over\partial \theta} +
  r{\partial\over\partial r}\left({\mean{u_\theta}\over
      r}\right)\right].
\label{laM}
\end{equation}
In the earlier definitions, no $\Omega$ factor was included
in the first term because,
on theoretical grounds, one expects the rotational effects on the
meridional part of the Reynolds stress $Q_{r\theta}$ to be proportional to
$\sin\theta\cos\theta$ \citep{R89,RBMD94}.
However, to obtain the same units for the coefficient $\LLM$ as for
the other components of $\Lambda$, we include here an $\Omega$ factor.
In \App{MerLamEffect}, we give a simplistic derivation for these
coefficients, which confirms that $Q_{r\theta}$ is not only
proportional to a higher power of $\Omega$ than the other two
components of $\Lambda$, but that it also picks up a contribution
proportional to $\mean{u}_\theta/r$.
However, under nearly isotropic conditions, the contribution of $\LLM$
should have the same sign as $\mean{u}_\theta$; see
\App{MerLamEffect}.
In our simulations this is indeed the case; see \Fig{Rey_mer}.

We can now solve \Eqss{laH}{laM} for $\LLM$, $\LLV$ and $\LLH$
\begin{equation}
\LLM={Q_{r\theta}\over \sin{\theta}\cos{\theta} \,\Omega}
+{\nut\over\sin{\theta}\cos{\theta} \,\Omega}\left[{1\over r}{\partial
    \mean{u_r}\over\partial \theta} + r{\partial\over\partial r}\left({\mean{u_\theta}\over
      r}\right)\right]\label{laM2}
\end{equation}
\begin{equation}
\LLV={Q_{r\phi}\over \sin{\theta}\,\Omega}+\nut r
\frac{\partial\ln\Omega}{\partial r},\label{laV2}
\end{equation}
\begin{equation}
\LLH={Q_{\theta\phi}\over \cos{\theta}\,\Omega}+\nut \tan{\theta}\,
\frac{\partial\ln\Omega}{\partial \theta}.
\label{laH2}
\end{equation}

We plot the Reynolds stresses ($Q_{r\theta}$, $Q_{r\phi}$, and
$Q_{\theta\phi}$), the turbulent viscosity $\nut$, and the three
components of the $\Lambda$ effect ($\LLM$,
$\LLV$ and $\LLH$) in the meridional plane for Runs~A1, A3t, and A5 in
\Fig{Rey_mer} and as a latitudinal cut (15$^\circ$ latitude) for all
runs of Set~A in \Fig{Rey_A}.
For Run~A1, the Reynolds stresses show the usual behavior of 
rotating convection \citep{KMGBC11,KKOWB16}.
$Q_{r\theta}$ and $Q_{\theta\phi}$ are antisymmetric over the equator.
In the northern hemisphere, $Q_{r\theta}$ is negative at the surface 
and positive deeper down, but only outside the inner tangent cylinder.
The latitudinal variation of $Q_{r\theta}$ agrees with that found
both by \cite{PTBNS93} and \cite{RBMD94}.
$Q_{\theta\phi}$ is positive in the northern hemisphere near the
surface and weakly negative deeper down at low latitudes.
As expected, $Q_{r\phi}$ is symmetric over the equator \citep{R80}.
This stress component is mostly positive (negative) at low (high)
latitudes with a small negative region at the equator.
The meridional structure of all Reynolds stress components agrees
qualitatively with Run~A6 of \cite{KMGBC11} and \cite{HRY15} with
significantly lower and higher density stratifications, respectively.
However, the peak values are half of those in Run~A6 of
\cite{KMGBC11}, which is likely due to the slower rotation in their study, as the
stresses are known to be quenched for faster rotation
\citep[e.g.,][]{R89,RKH13}.

The turbulent viscosity $\nut$ has a maximum at the equator and
towards the bottom of the convection zone.
The value of $\nut/\nu$ is close to Reynolds number $\Rey=35$.
$\LLM$ is non-zero only at low latitudes, being positive near the
surface in the topmost 5--10 Mm, negative a bit deeper and mostly positive with a lower
amplitude even deeper down.
$\LLV$ shows strong alignment with the rotation axis, being positive
(negative) outside (inside) a cylindrical radius of $0.85\,R$, with a
negative region near the surface at the equator.
$\LLH$ has a similar structure, but shows a concentration near the
equator above $0.85\,R$, where $\LLH$ is three times stronger than $\LLV$.
A similar structure, but with three times lower values, has been found by
\cite{KMB14} and \cite{KKKBOP15} using the same technique, but for
runs that rotate slower.

The presence of a coronal envelope significantly alters the Reynolds
stresses and therefore $\nut$ and the $\Lambda$ effect; see
\Figs{Rey_mer}{Rey_A}.
The Reynolds stresses lose their alignment with the rotation axis.
$Q_{r\theta}$ changes sign and is positive (negative) at lower
latitudes in the northern (southern) hemisphere, which is similar
to \cite{HRY15}.
This behavior has been found for all runs with an extended coronal envelope
(A2, A3, A3t, A4, and A5), whereas in Runs~A1, A1c, and A1c2, $Q_{r\theta}$
is negative close to the surface in the northern hemisphere; see
\Fig{Rey_A}(a).
This seems to confirm the presence of a correlation between a positive
(negative) value of
$Q_{r\theta}$ in the northern (southern) hemisphere and the generation
of near-surface negative shear.
The inclusion of a coronal layer changes $Q_{r\phi}$ such that the
minimum at the equator disappears and the overall magnitude of the
stress is reduced by roughly a factor of two; see \Fig{Rey_mer}.
At 15$^\circ$ latitude, all runs show a similar behavior with the main
variation being the location of the maximum near the surface depending
on the depth of the cooling layer; see \Fig{Rey_A}(b).
$Q_{\theta\phi}$ changes similarly as $Q_{r\phi}$.
The maxima near the surface are again shifted deeper in the runs with
a deeper acting cooling layer; see \Fig{Rey_A}(c).
The profiles of $Q_{r\phi}$ and $Q_{\theta\phi}$ of Run~A3t are
similar to \cite{HRY15}.

The turbulent viscosity $\nut$ varies less as a function latitude
in runs with a coronal envelope; see \Fig{Rey_mer}.
Figure~\ref{Rey_A}(d) shows that $\nut$ is reduced at the bottom of
the convection zone and near the surface in the runs with a coronal
envelope which is consistent with the lower turbulent velocities in
those cases; see \Fig{strat2_A}(a).
In Run~A5, the profile is similar to the other runs with coronal
envelope, but $\nut/\nu$ is higher due to a lower value of $\nu$.

All of the $\Lambda$-coefficients are reduced in runs with a coronal
envelope; see \Figs{Rey_mer}{Rey_A}(e)--(g).
This is consistent with a weaker differential rotation in
these runs; see \Fig{diff_cut}.
Similarly as $Q_{r\theta}$, also $\LLM$ changes sign in the equatorial
regions in the runs with a coronal envelope. This is consistent with
the mean meridional flows changing sign in these runs as can be seen
in \Fig{Rey_A}(h).
$\LLV$ does not change significantly apart from the reduced amplitude.
The vertical $\Lambda$ effect is thought to be responsible for
generating radial shear. This is consistent with a stronger $\LLV$
producing a stronger radial differential rotation in Run~A1 and with a
correspondingly weaker $\LLV$ and weaker radial differential rotation
in Runs~A2, A3, A3t, A4, and A5; see \Fig{diff_cut}.
The horizontal $\Lambda$ effect shows a more broad maximum as a
function of latitude in the near surface layers in Run~A3t in
comparison to Run~A1, whereas the amplitude is reduced by a factor of
four.
Figure~\ref{Rey_A} reveals that the values of $\LLH$ near the surface are
lower in the runs with a coronal envelope or a cooling function
penetrating the surface, consistent with weaker latitudinal
shear.
The changes due to the coronal envelope are even more pronounced in
the anisotropy parameters $A_{\rm M}$, $A_{\rm V}$, $A_{\rm H}$, which
are related to the $\Lambda$ effect; see
\App{aniso} and \Fig{An_mer}.
$A_{\rm M}$ and $A_{\rm V}$ change from being mostly positive to
mostly negative (also within the coronal layer), whereas the changes
in $A_{\rm H}$ are not as strong.
The results for the more turbulent Run~A5 are very similar with those
of Run~A3t; see \Fig{Rey_mer}.

In Set~B, where rotation is slower, the rotational influence on
convection is reduced by roughly a factor of two.  However, the only
major difference to the runs in Set~A occurs for $Q_{r\theta}$, where
the sign at low latitudes near the surface is positive already without
a coronal envelope, although the difference in the meridional
$\Lambda$ effect is rather minor. The main effect of the coronal
envelope on the Reynolds stresses as well as the components of
the $\Lambda$ effect is their reduced amplitude similarly as in Set~A.

According to theory \citep[e.g.,][]{KR95,KR05,RKT14,K16}, the near-surface
shear layer in the Sun is caused by the vanishing horizontal $\Lambda$
effect and the sole contribution of a negative vertical $\Lambda$ effect.
In our simulations, we do indeed find a weaker $\LLH$ in some of the runs, where
$\dd \ln \Omega/\dd \ln r$ is negative, see \Figs{Rey_mer}{Rey_A}(g),
but for most of the runs the relation is inconclusive.
In \Sec{sec:therm}, we will investigate this in more detail.

In this section we have not discussed the influence of the magnetic field
on differential rotation generators.
This has been discussed in detail in \cite{KKKBOP15} for
similar simulations, but without coronal envelope.
They found that the large-scale and turbulent Maxwell stresses
are around two orders of magnitude smaller than the Reynolds stresses
and therefore do not influence the angular momentum transport
significantly.
However, the recent study by \cite{KKOWB16} has revealed that
the turbulent Maxwell stresses in similar setups without coronal
envelopes are of the same order of magnitude as the Reynolds stresses
at high magnetic Reynolds numbers. The current runs are in the
intermediate range in $\Rm$ where the total stress is already affected
but the qualitative character of the hydrodynamic results remains
unchanged.

\subsection{Thermal wind balance}
\label{sec:therm}
\begin{figure}[t!]
\begin{center}
\includegraphics[width=\columnwidth]{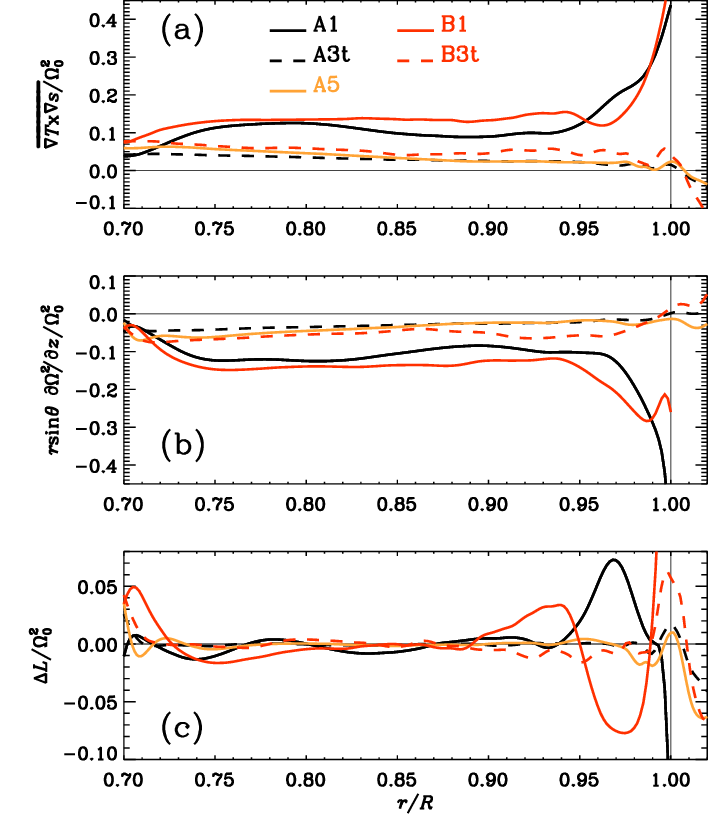}
\end{center}\caption[]{
The two dominant terms of \Eq{eq:baroc} and their difference $\Delta
L=r\sin{\theta}{\partial\Omega^2/\partial z} +\left(\overline{\nab
    T\times \nab s}\right)_\phi$ for Runs~A1, A3t, A5, B1, and B3t in the
northern hemisphere at 15$^\circ$ latitude.
The thin black lines indicate the zero value and the surface ($r=R$).
}
\label{bar}
\end{figure}

The Taylor-Proudman balance can be broken by a non-zero baroclinic
term in the mean azimuthal vorticity $\mean{\omega_\phi}$ equation,
also know as the thermal wind balance or meridional circulation
evolution equation \citep[e.g.,][]{BMT92b,KR95,WKMB13},
\begin{equation}
{\partial\mean{\omega_\phi}\over\partial
  t}=r\sin{\theta}{\partial\Omega^2\over\partial z} 
  +\left[\overline{\nab T\times \nab s}\right]_\phi -
  \left[\nab\times\left({1\over\meanrho}\,\nab\cdot\meanrho\,
\overline{\fluc{\uu}\fluc{\uu}}\right)\right]_\phi\!,
 \label{eq:baroc}
\end{equation}
where $\partial/\partial z=\cos\theta\,\partial/\partial r
-r^{-1}\!\sin\theta\,\partial/\partial\theta$ is the derivative along
the rotation axis and $\mean{\oo}=\nab\times\meanuu$ is the mean
vorticity. 
We neglect here the contribution of the Maxwell stress
arising from the correlations of the fluctuating magnetic field and
those from the mean magnetic field.
The contributions of meridional flows turn out to be small and can be
neglected.
\cite{WKMB13} could show that the baroclinic term, that is
the second term in \Eq{eq:baroc}, is responsible for spoke-like
differential rotation profile.
Following the study of \cite{WKMB13}, we plot the two dominant terms
on the right hand side of \Eq{eq:baroc} together with the residual
$\Delta L= r\sin{\theta}{\partial\Omega^2/\partial z} +
\left[\overline{\nab T\times \nab s}\right]_\phi$ in \Fig{bar} for
Runs~A1, A3t, A5, B1, and B3t.
The meridional distributions of $\left[\overline{\nab T\times \nab
    s}\right]_\phi$ and $r\sin{\theta}{\partial\Omega^2/\partial z}$
of Runs~A1 and Runs~A3t are very similar to Figures 5 and 9 of
\cite{WKMB13}, respectively.
For most of the convection zone the two terms match well and $\Delta
L$ is close to zero.
In the upper $0.1\,R$, $\Delta L$ is non-zero.
In the case of Run~A1, $\Delta L$ is positive in the region above
$r=0.94\,R$.
However, right at the surface, it is negative, probably because of
boundary effects.
By contrast, Run~B1 develops regions, where $\Delta L$ is positive
($0.90\,R\le r\le 0.94\,R$) and where $\Delta L$ is even more strongly
negative ($0.95\,R\le r\le 0.99\,R$).
In Runs~A3t, A5, and B3t, the overall magnitude of terms in the thermal wind
balance is lower, resulting in a significantly smaller residual
$\Delta L$.
However, $\Delta L$ seems to be close to zero in all three runs and becomes
non-zero only near the surface; see \Fig{bar}.
Thus, there is no clear change in behavior depending on rotational
influence or Prandtl numbers in the coronal runs.
As the meridional circulation is stationary, apart from cycle dependent
variations, the left-hand side of \Eq{eq:baroc} is on average zero.
Therefore, in the region close to the surface the contribution of the Reynolds stress,
that is the third term on the rhs of \Eq{eq:baroc}, plays a more
important role and will be related to the residual $\Delta L$.
We can rewrite this term
\begin{eqnarray}
\Delta L&\approx&\left[\nab\times\left({1\over\meanrho}\,\nab\cdot\meanrho\,\overline{\fluc{\uu}\fluc{\uu}}
  \right)\right]_\phi
=
-{1\over\meanrho^2}\left[\nab\meanrho\times\nab\cdot\meanrho\,\overline{\fluc{\uu}\fluc{\uu}}\right]_\phi\nonumber\\
&=&
-{1\over\meanrho^2}\left(\partial_r\meanrho\left[\nab\cdot\meanrho\,\overline{\fluc{\uu}\fluc{\uu}}\right]_\theta
-{1\over
  r}\partial_\theta\meanrho\left[\nab\cdot\meanrho\,\overline{\fluc{\uu}\fluc{\uu}}\right]_r\right).
\end{eqnarray}
The latitudinal component of the divergence of
$\meanrho\,\overline{\fluc{\uu}\fluc{\uu}}$ is given by
\begin{eqnarray}
\left[\nab\cdot\meanrho\,\overline{\fluc{\uu}\fluc{\uu}}\right]_\theta
&=&{1\over r^2} \partial_r\left(r^2\meanrho\,\overline{\fluc{u_r}\fluc{u_\theta}}\right) +
  {\meanrho\,\overline{\fluc{u_r}\fluc{u_\theta}}\over r}\nonumber\\
&+& {1\over r\sin\theta} \partial_\theta
\left(r\sin\theta\,\meanrho\,\overline{\fluc{u_\theta}
    \fluc{u_\theta}}\right)-
{\cot\theta\,\meanrho\,\overline{\fluc{u_\phi}\fluc{u_\phi}}\over r}
\label{eq:qrt}
\end{eqnarray}
and the radial component of the divergence is given by
\begin{eqnarray}
\left[\nab\cdot\meanrho\,\overline{\fluc{\uu}\fluc{\uu}}\right]_r
&=&{1\over r^2} \partial_r\left(r^2\meanrho\,\overline{\fluc{u_r}\fluc{u_r}}\right) +
 {1\over r\sin\theta} \partial_\theta
\left(r\sin\theta\,\meanrho\,\overline{\fluc{u_\theta}
    \fluc{u_\theta}}\right)\nonumber\\
&-&{\meanrho\,\overline{\fluc{u_\theta}\fluc{u_\theta}}+\meanrho\,\overline{\fluc{u_\phi}\fluc{u_\phi}}\over
  r},
\label{eq:qrr}
\end{eqnarray}
where
$\overline{\fluc{u_i}\fluc{u_j}}=\overline{\fluc{u_j}\fluc{u_i}}$ and
${\partial/\partial\phi}=0$ because of the azimuthal mean.
This means that the contributions related to
$\overline{\fluc{u_r}\fluc{u_\phi}}$, and
$\overline{\fluc{u_\theta}\fluc{u_\phi}}$ are zero.
Furthermore, we find that terms related to
$\overline{\fluc{u_\phi}\fluc{u_\phi}}$ are too small to have a
strong effect.
\begin{figure}[t!]
\begin{center}
\includegraphics[width=\columnwidth]{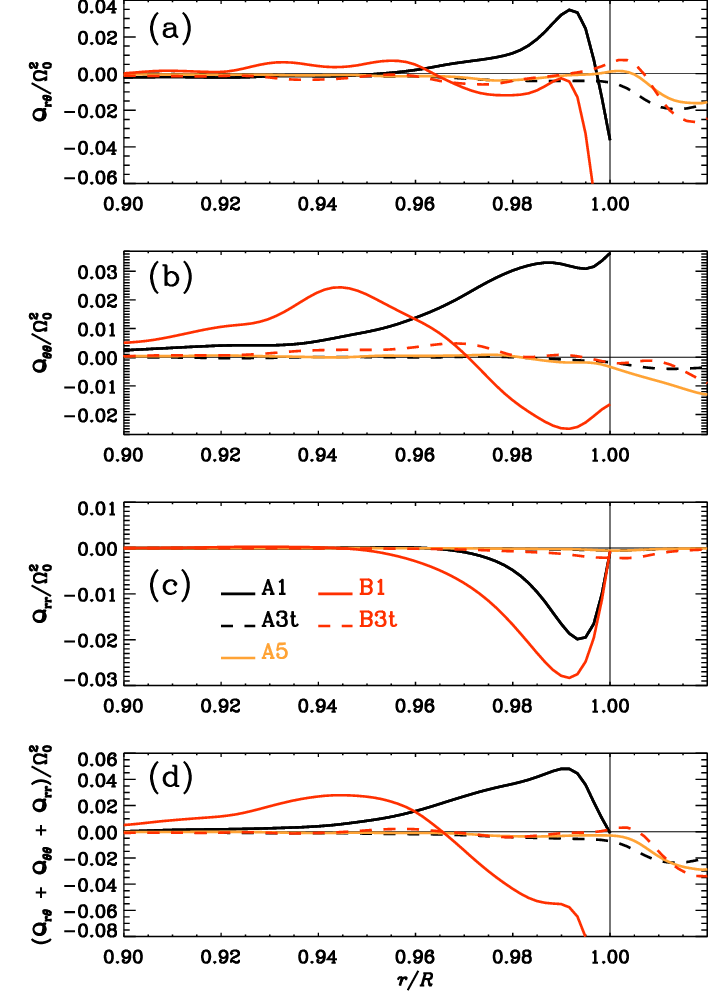}
\end{center}\caption[]{
The three dominant contributions of the Reynolds stress ${\cal
  Q}_{r\theta}$ (\ref{qqrt}), ${\cal Q}_{\theta\theta}$ (\ref{qqtt})
and ${\cal Q}_{rr}$ (\ref{qqrr}) as well as their sum for Runs~A1,
A3t, A5, B1, and B3t in the northern hemisphere at 15$^\circ$ latitude.
The thin black lines indicate the zero value and the surface ($r=R$).
}
\label{barII}
\end{figure}
In \Eq{eq:qrt} the first three terms and in \Eq{eq:qrr} only the first
term have a strong contribution to the thermal wind balance.
We summarize them in the following three expressions:
\begin{equation}
-{1\over\meanrho^2}\partial_r\meanrho\left({1\over r^2} \partial_r\left(r^2\meanrho\overline{\fluc{u_r}\fluc{u_\theta}}\right) +
  {\meanrho\,\overline{\fluc{u_r}\fluc{u_\theta}}\over
    r}\right)\equiv{\cal Q}_{r\theta},
\label{qqrt}
\end{equation}
where the first term on the lhs is the dominant one,
\begin{equation}
-{1\over\meanrho^2}\partial_r\meanrho\left({1\over r\sin\theta} \partial_\theta
\left(r\sin\theta\,\meanrho\,\overline{\fluc{u_\theta}\fluc{u_\theta}}\right)\right)
\equiv{\cal Q}_{\theta\theta},
\label{qqtt}
\end{equation}
and
\begin{equation}
+{1\over\meanrho^2}{1\over r}\partial_\theta\meanrho\left({1\over
    r^2} \partial_r
  \left(r^2\meanrho\overline{\fluc{u_r}\fluc{u_r}}\right)\right) \equiv{\cal Q}_{rr},
\label{qqrr}
\end{equation}
In \Fig{barII}, we plot all three terms and their sum for the same runs as
in \Fig{bar}.
In Run~A1, their sum is positive, resulting in a negative contribution
of the Reynolds stresses to the thermal wind balance; see \Eq{eq:baroc}.
In Run~B1, these terms are positive around $r=0.94\, R$ and negative
closer to the surface.
${\cal Q}_{rr}$ is in all runs negative giving a positive contribution
to the thermal wind balance.
The runs with a coronal envelope seem to have a weaker and negative
contribution from the Reynolds stresses.
This is probably related to the low density stratification near the
surface.
If we compare the plots of \Fig{barII} with \Fig{bar}, there seems
to be some agreement of the residual $\Delta L$ with the contributions
of the Reynolds stresses ${\cal Q}_{r\theta}$, ${\cal
  Q}_{\theta\theta}$, and ${\cal Q}_{rr}$.
However, there seems to be a shift in radius, which might be related
to still missing contributions, which we could not identify in 
the present work.
This might be related to the fact that in our simulations, in particular
near the surface, $\nab\cdot\meanuu$ is not zero as assumed in many
mean-field models \citep[e.g.,][]{KR95,KR05}.
Furthermore, we have here neglected the contribution of the Maxwell
stresses, which might also explain this inconsistency.
However, we can confirm the result from \cite{HRY15}
that $Q_{r\theta}$ is important near the surface and gives an
important contribution to the thermal wind balance near the top boundary.
However, this contribution is not directly related to the stress
$Q_{r\theta}$ itself but to its radial gradient.
Furthermore, also $Q_{\theta\theta}$ and $Q_{rr}$ have a large
contribution to the
thermal wind balance in terms of ${\cal Q}_{\theta\theta}$ and ${\cal
  Q}_{rr}$.
Similar to $Q_{r\theta}$ in \Fig{Rey_A}, ${\cal Q}_{r\theta}$ changes
sign in the runs where we have identified a near-surface shear layer;
see \Sec{sec:diff}.
This effect is even stronger near the equator, where it could play
a role in generating the near-surface shear layer.

\subsection{Radial field boundary condition}
\label{sec:radfield}

\begin{figure}[t!]
\begin{center}
\includegraphics[width=0.24\textwidth]{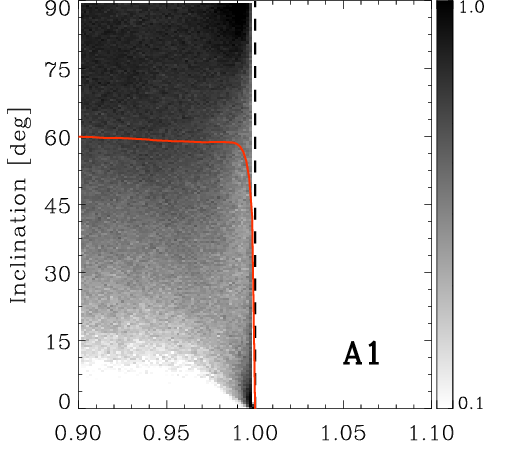}
\includegraphics[width=0.24\textwidth]{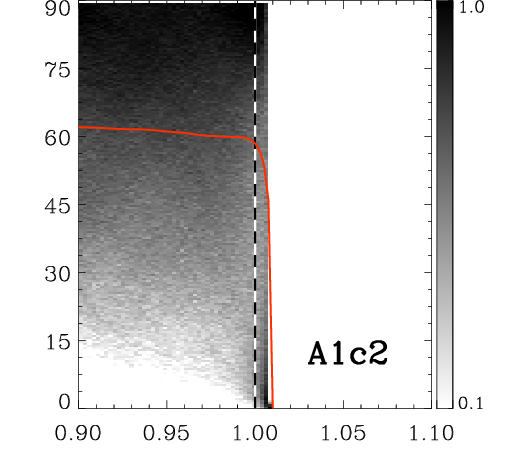}
\includegraphics[width=0.24\textwidth]{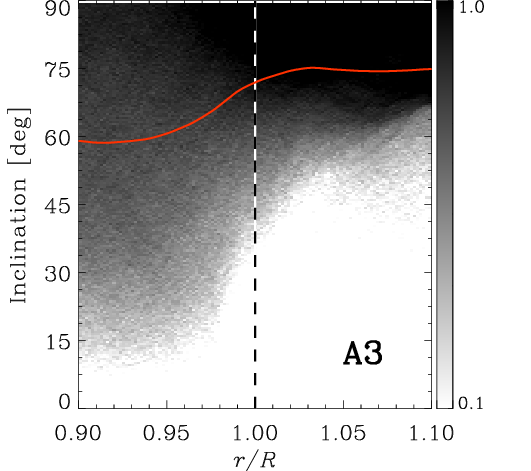}
\includegraphics[width=0.24\textwidth]{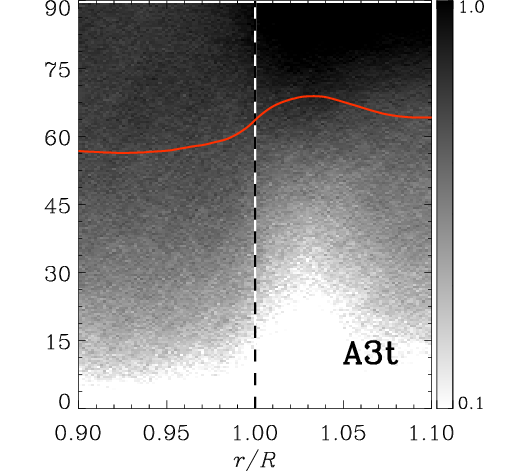}
\end{center}\caption[]{
Inclination of the magnetic field near the surface plotted as a 2D
histogram over radius $r/R$ for Runs~A1, A1c2, A3, and A3t.
0$^\circ$ means fully radial and 90$^\circ$ fully horizontal.
The red line indicates the average for each radii.
}
\label{rad2_A}
\end{figure}

In the case of Runs~A1 and B1, we employ a radial field condition at
the surface, $r=R$, whereas in the other cases the radial condition is
enforced either at $r=1.01\,R$ (Runs~A1c, A1c2 and B1) or high above
the convection zone (Runs~A2, A3, A3t, A4, A5, B3, B3t).
This affects the magnetic field distribution and therefore the dynamo
in the convection zone.
In the majority of the upper convection zone of Run~A1, the angle
between
the field and the radial direction is distributed nearly uniformly, as
indicated by a mean of the absolute angle being close to $60^\circ$.
The field is fully radial just in the two uppermost grid points;
see \Fig{rad2_A}.
In Runs~A1c and A1c2 the field distribution is the same, except that
the magnetic boundary condition is placed higher.
For the runs with coronal envelope, the field is less radial at the
surface than in Run~A1.
Furthermore, the field is more horizontal in the coronal envelope than
in the convection zone.
The comparison of Runs~A3 and A3t suggests that the cooling layer,
which reaches deeper in Run~A3, allows the field to become less radial
at and above the surface.
This leads us to conclude that the radial boundary condition does not
significantly affect the structure and inclination below $r=0.98\,R$.

\cite{WRKKB16} find turbulent radial downward pumping from a similar
run, which can cause radial 
alignment of the magnetic field, but only close to the surface
($r>0.98\,R$) and at low latitudes,
which is consistent with our findings.
In local convection simulations, it has been found that turbulent
pumping is a major effect causing horizontal magnetic field to move
downward \citep[e.g.,][]{NBJRRST92,TBCT98,OSBR02}.
It is also believed that this effect can cause the field near the
surface to become dominantly radial \citep{CSJI12} and is therefore
the main reason for the success of surface flux transport models
\citep[see][for a review]{MY:2012}.

\subsection{Cyclic dynamo solutions with dynamo wave propagation}
\label{sec:dynamo}

\begin{figure*}[t!]
\begin{center}
\includegraphics[width=0.196\textwidth]{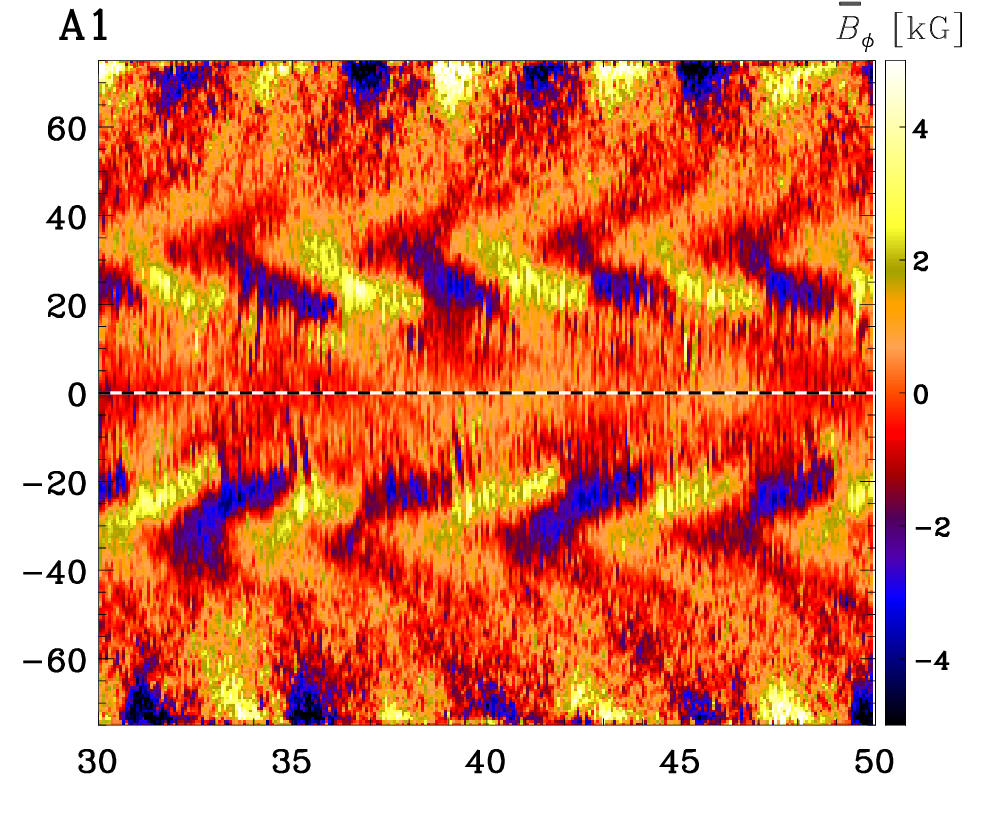}
\includegraphics[width=0.196\textwidth]{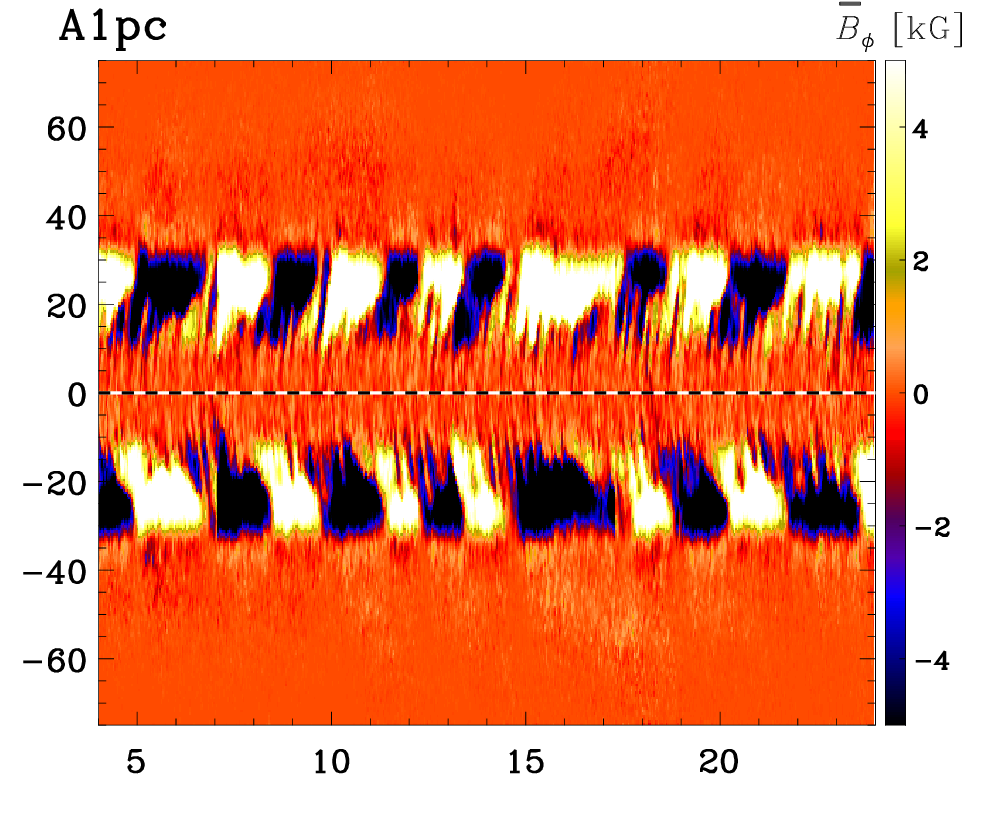}
\includegraphics[width=0.196\textwidth]{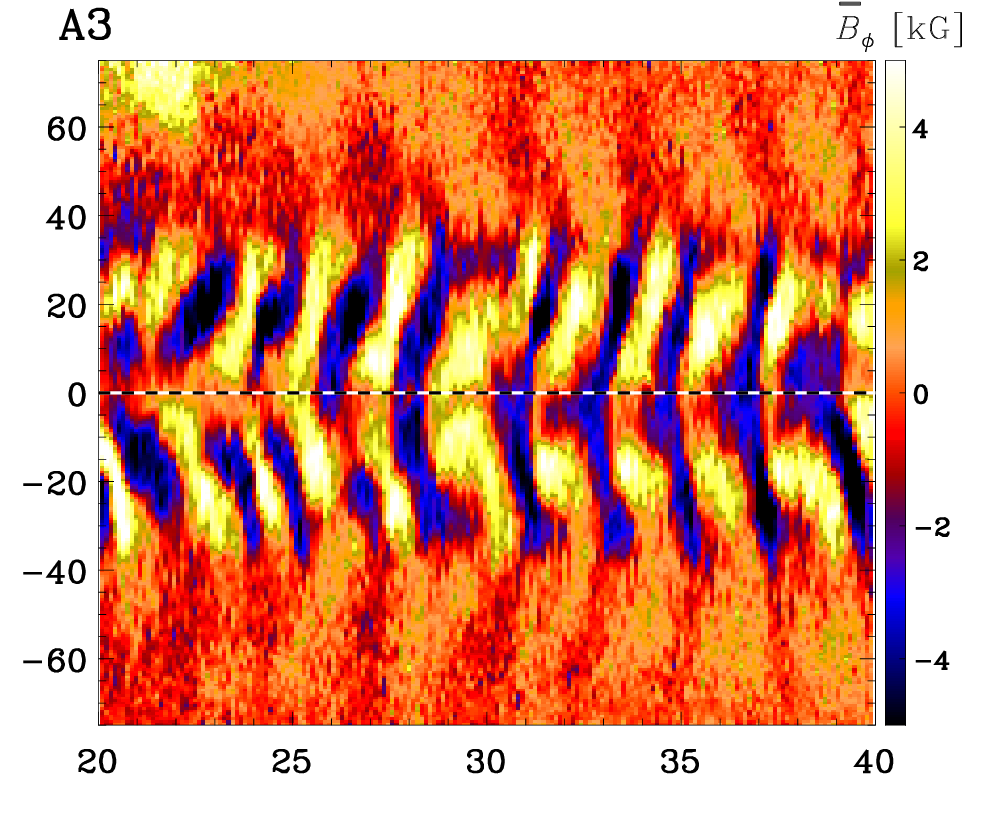}
\includegraphics[width=0.196\textwidth]{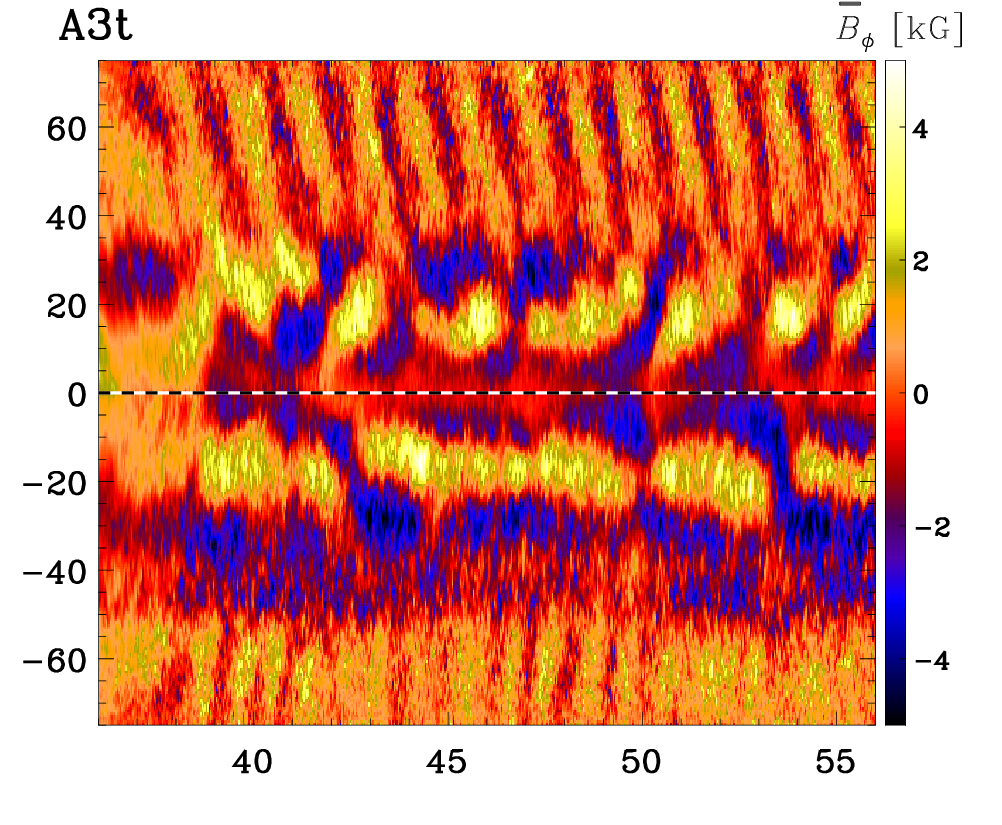}
\includegraphics[width=0.196\textwidth]{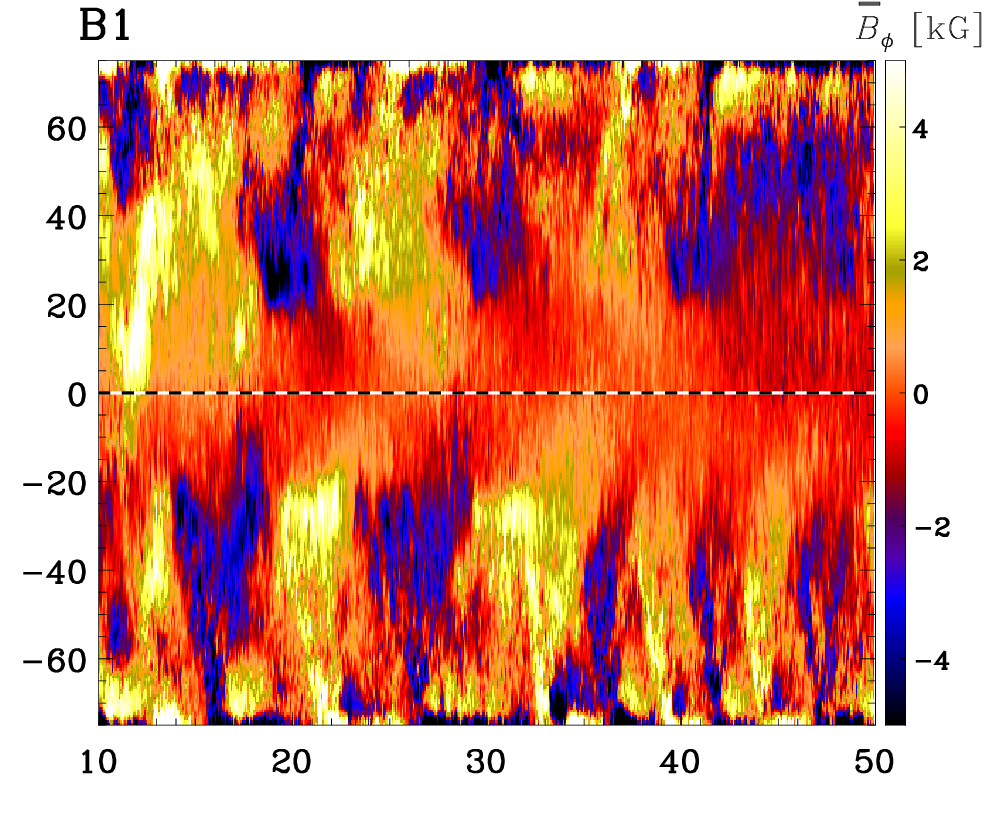}
\includegraphics[width=0.196\textwidth]{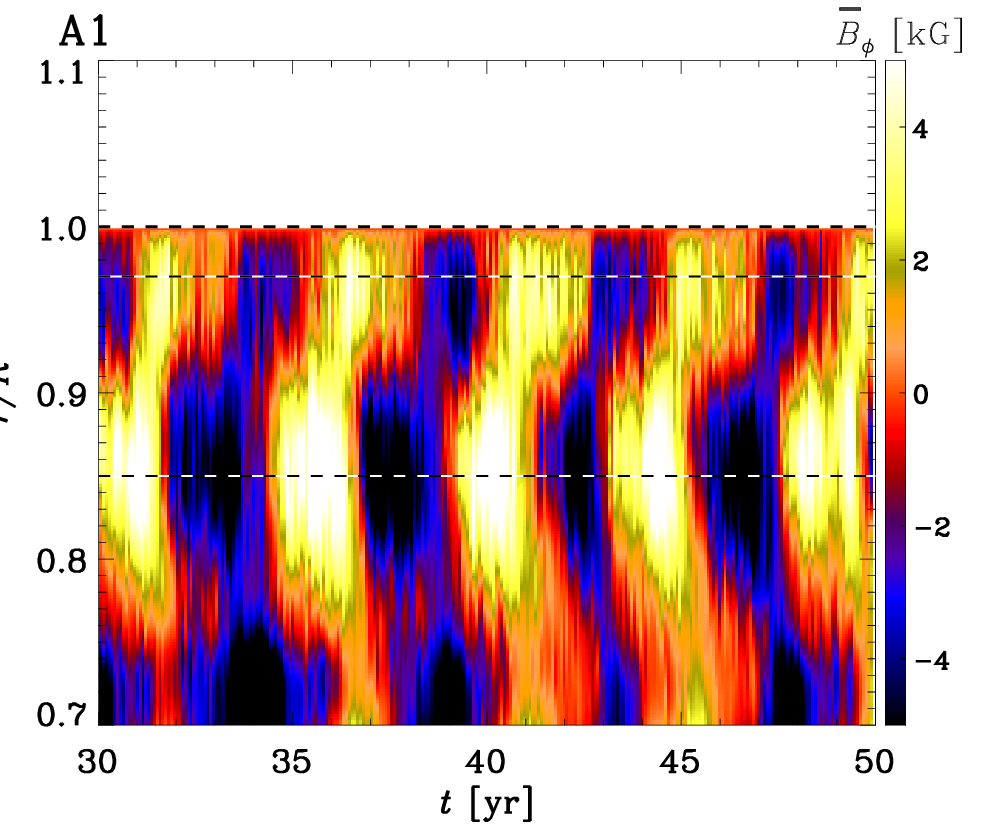}
\includegraphics[width=0.196\textwidth]{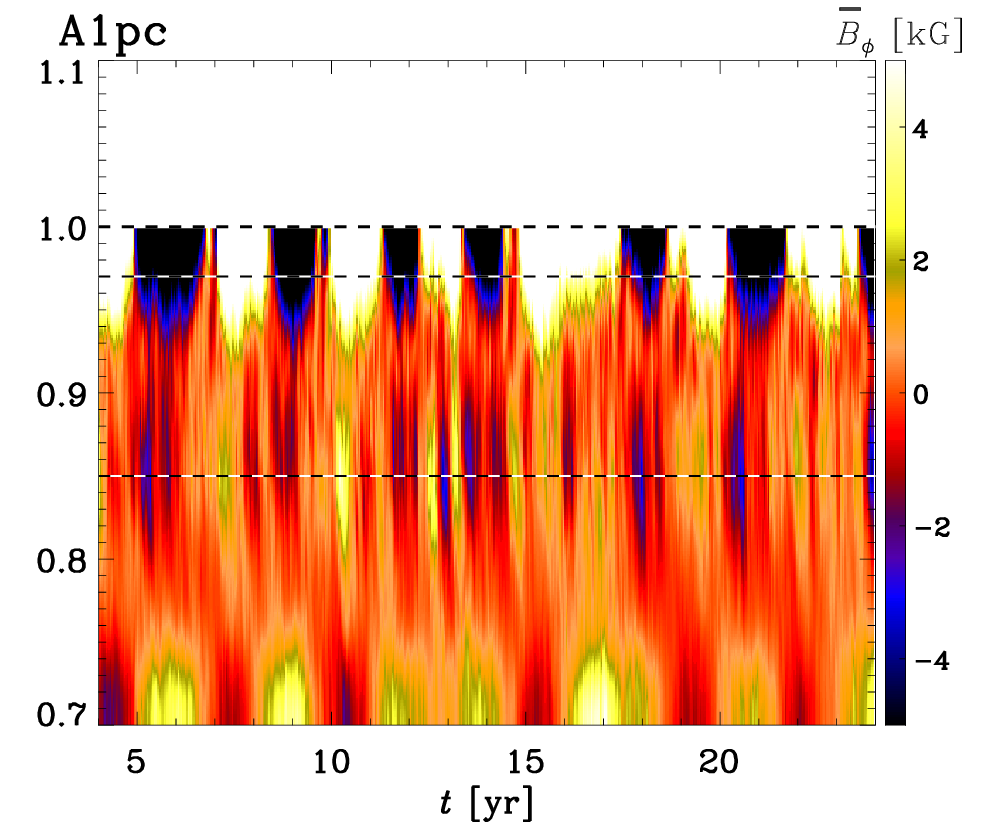}
\includegraphics[width=0.196\textwidth]{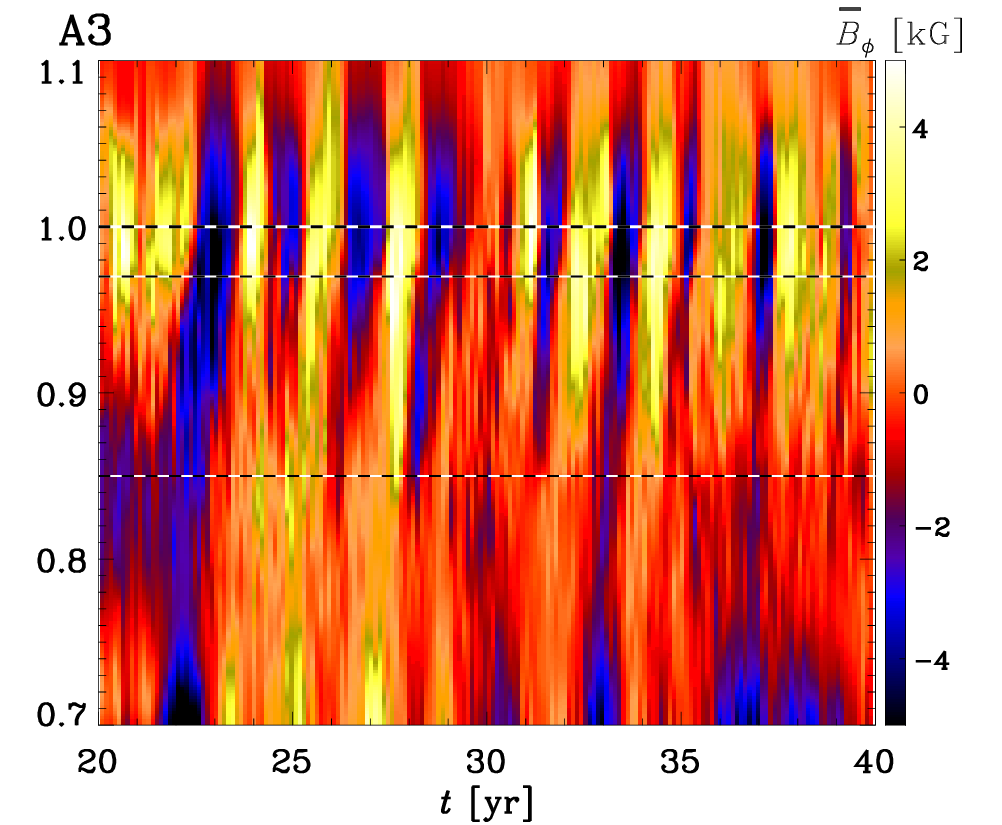}
\includegraphics[width=0.196\textwidth]{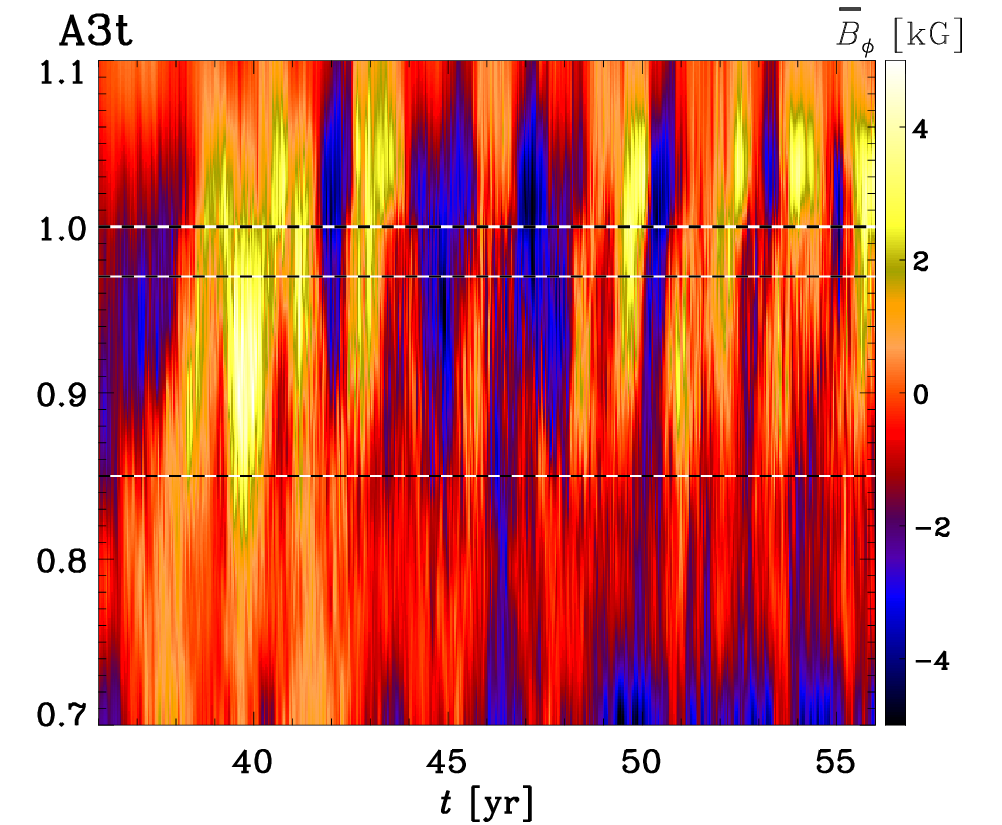}
\includegraphics[width=0.196\textwidth]{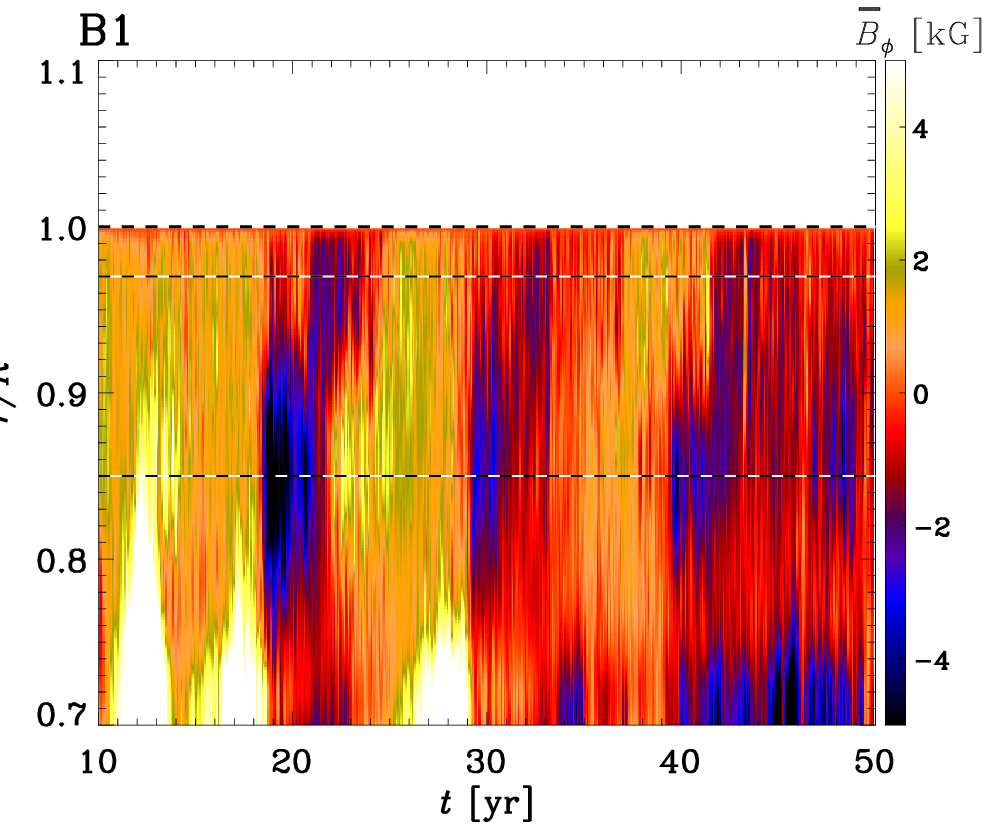}
\includegraphics[width=0.196\textwidth]{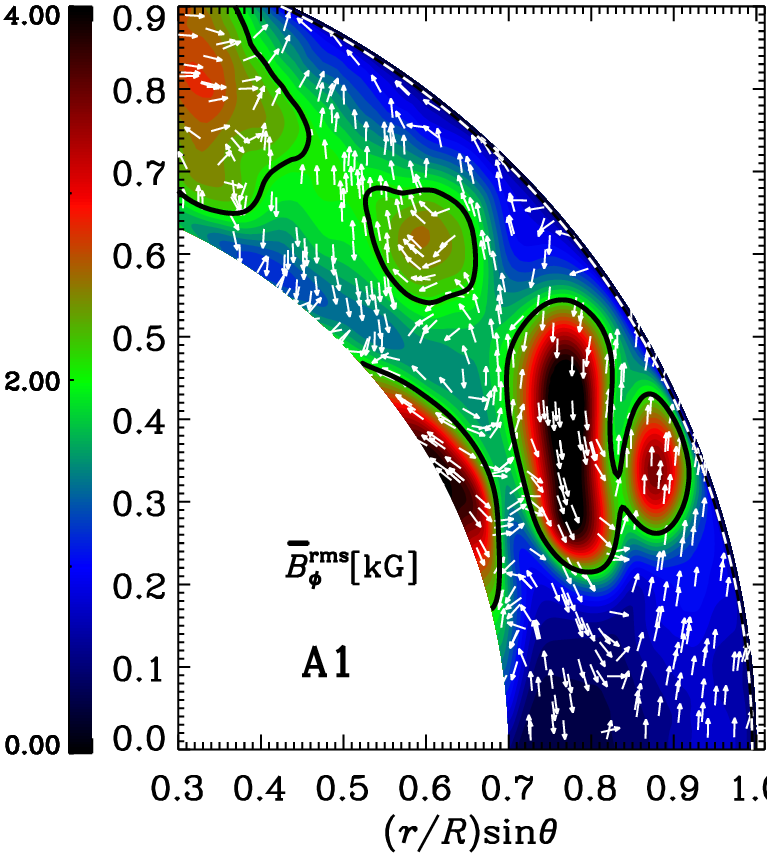}
\includegraphics[width=0.196\textwidth]{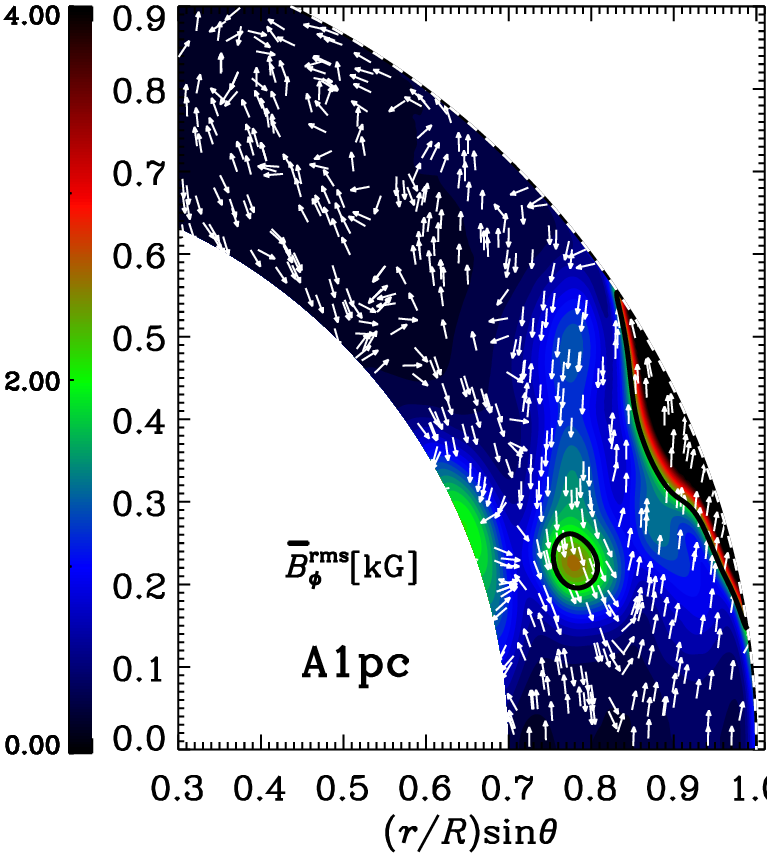}
\includegraphics[width=0.196\textwidth]{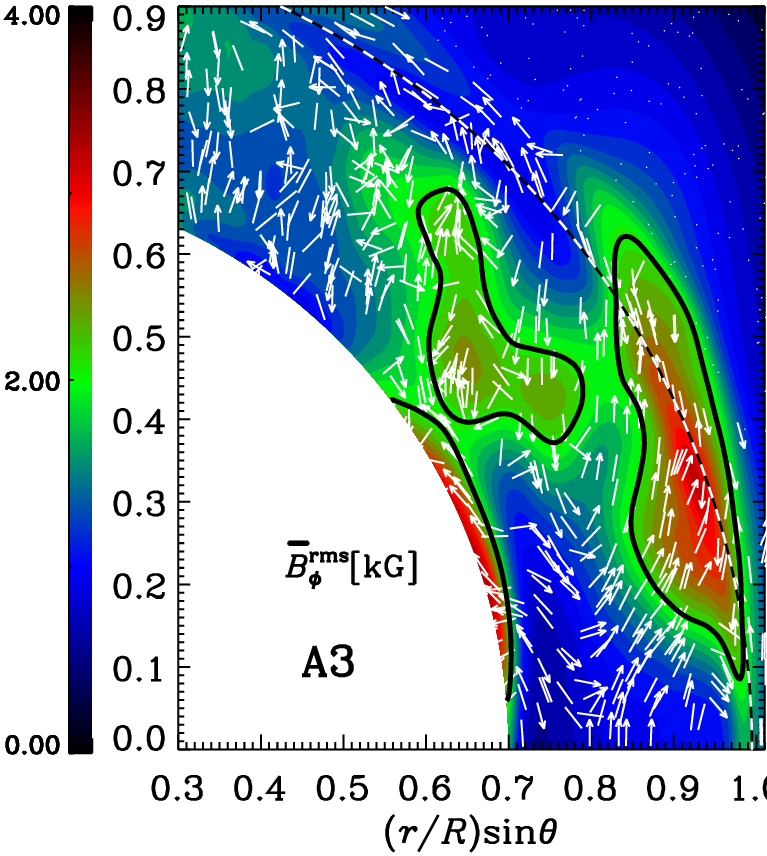}
\includegraphics[width=0.196\textwidth]{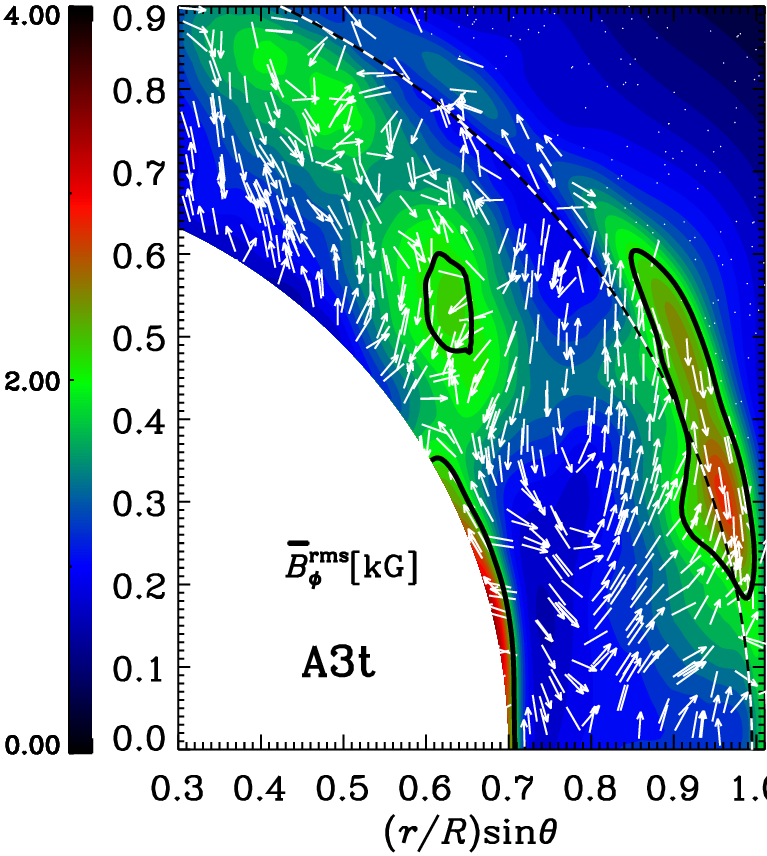}
\includegraphics[width=0.196\textwidth]{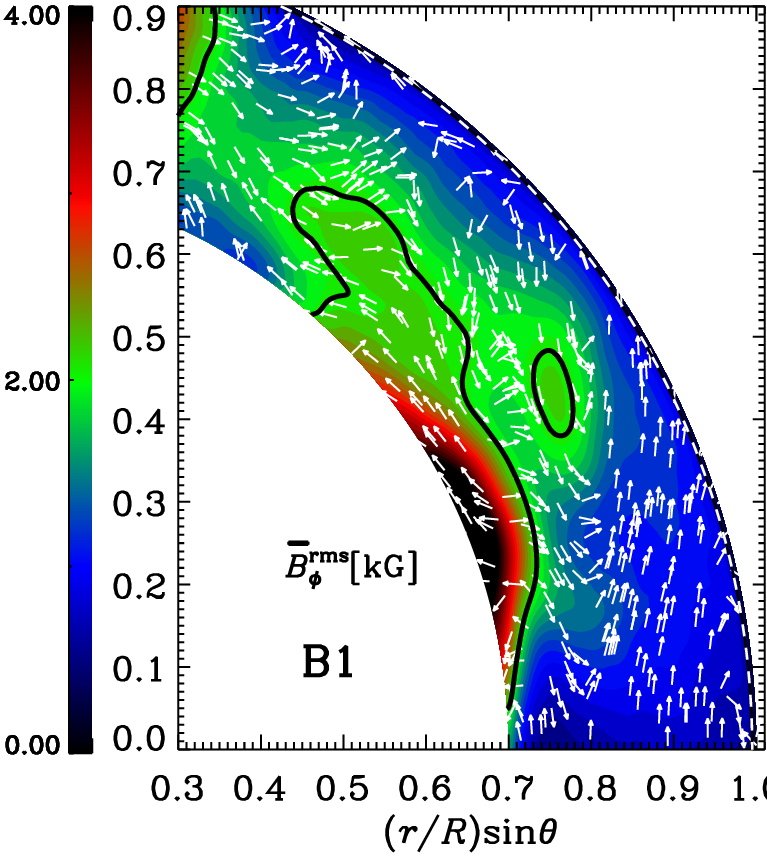}
\end{center}\caption[]{
Mean toroidal magnetic field $\mean{B_\phi}$ evolution as
time-latitude (butterfly) diagram, plotted at a radius $r=0.98\,R$
({\it top row}) and time-radius diagram plotted at a 25$^\circ$
latitude ({\it middle row}) in kG during a $20\yr$ interval in the
saturated stage for Runs~A1, A1pc, A3, A3t, and B1.
For Run~B1, we plot the butterfly diagram at radius
$r=0.84\,R$ and its time interval is 40 yr.
The dashed horizontal lines mark the equator ($\theta=\pi/2$) and
the radii $r=R$, $r=0.98\,R$ and $r=0.85\,R$, respectively.
Propagation of the mean magnetic field for the same
runs ({\it bottom row}).
Color coded $\Btt$ is plotted during the saturated stage together with
white arrows showing the direction of migration $\ssss_{\rm
  mig}(r,\theta)=-\alpha \eee_\phi \times\nab\Omega$ of an
$\alpha\Omega$ dynamo wave \citep{P55,Yos75}; see \cite{WKKB14}.
We suppress the arrows above $r=1.05$.
The black solid lines indicate isocontours of $\mean{B_\phi}$ at 2.0\,kG.
The dashed white lines indicate the surface ($r=R$).
}
\label{butA}
\end{figure*}

In all runs of both sets, convective motions, overall rotation, and
their interaction contribute to generating a large-scale magnetic field.
Most of the runs produce cyclic magnetic fields in the saturated
phase.
In \Fig{butA}, we plot the time evolution of the mean toroidal
magnetic field $\mean{B_\phi}$ for a selection of runs.
To investigate the cause of the propagation direction of the magnetic
field, we apply the same technique as in \cite{WKKB14} and
\cite{KKOBWKP16} and we calculate the propagation direction using the so-called
Parker--Yoshimura rule \citep{P55, Yos75}.

As discussed in \cite{WKKB14} in detail, Runs~A1 and A1c2 show a
solar-like equatorward migration of the mean field.
This is probably caused by an equatorward propagating dynamo wave
generated by the region of negative shear in the middle of convection
zone; see \Fig{diff} and bottom row of \Fig{butA}.
We have confirmed this interpretation using all turbulent transport
coefficients obtained with the test-field method \citep{WRKKB16}.
To reiterate, Runs~A1 and A1c2 have two dominant dynamo modes;
equatorward and poleward migrating branches at mid-latitudes and high
latitudes, respectively; see top row of \Fig{butA}.
The dynamo wave seems to be formed in the middle of the convection zone and propagates
toward the bottom and top of the convection zone; see bottom row of \Fig{butA}.
The cycle period is around five years \citep{WKKB14}.
Near the surface there is a weaker mode, which has a cycle period of
1.5 years that propagates poleward at mid-latitudes and vanishes at
higher latitudes.
At the bottom of the convection zone exists another dynamo
mode, see \Fig{butA}, which might cause a long-term variability.
Such competing dynamo modes have been analyzed in detail in \cite{KKOBWKP16}.
Furthermore, the magnetic field in Run~A1c shows no equatorward migration.
As described by \cite{WKKB14}, this is most likely caused by the much weaker
negative radial gradient of $\Omega$ in the middle of convection zone, also
visible in \Fig{diff_cut}(a) and the suppression of turbulent
convective motion near the surface, as discussed in \Sec{sec:flux}.
In this paper, we focus on the differences between the runs with and
without a coronal envelope.

Before we discuss the runs with a coronal envelope, we investigate how
the magnetic evolution changes, if we change the radial field condition
to a perfect conductor condition at surface (Run~A1pc).
This is sufficient to cause the dynamo modes to change.
Close to the surface the toroidal magnetic field becomes the
strongest, reaching values of more than 10 kG which are more than
two times higher than the maximum values in Runs~A1 and A1c2; see
top and middle row of \Fig{butA}.
There the predicted migration direction is poleward, which agrees
with the actual migration near the surface; see bottom row of
\Fig{butA}.
Furthermore, we find indication of a poleward migrating subdominant
dynamo mode similar as in Runs~A1, A1c, and A1c2.
However, we find an equatorward propagating mode in the middle of the
convection zone, which is most likely related to the smaller and weaker
concentration of $\Btt$ found in the middle of the convection zone
which is predicted to propagate equatorward.
It seems that the hydro-thermal setup tends to produce strong
toroidal magnetic fields near the surface, but the radial field
condition at the top boundary of Run~A1, A1c, and A1c2 prevents this.
The toroidal magnetic field is concentrated at mid-latitudes and
produces nearly no field above 35$^\circ$ latitude.
This is very surprising, given that the other runs without a coronal
envelope produce strong magnetic fields at high latitudes.
This absence of polar field in Run~A1pc suggests a relation
between the polar field and the radial field at the surface.
The relation might be interesting in view of the Babcock-Leighton dynamo
framework \citep[e.g.,][]{B61,L64,DC99}.

Adding a coronal envelope on top of the convection zone changes the
magnetic field evolution significantly; see \Fig{butA}.
For the runs of Set~A, the mean toroidal magnetic field migrates
mostly poleward in a region between the equator and $\pm40^\circ$
latitude.
At high latitudes the magnetic field is weaker, but shows a tendency of
equatorward migration, in particular in Run~A3t this equatorward
propagation is clearly visible.
In the middle of convection zone, the equatorward propagation also
dominates at lower latitudes.
As a main difference compared to Run~A1, the mean toroidal field
occurs close to the surface instead of in bulk of the convection zone.
At this location, the radial shear is positive causing a dynamo wave
to propagate poleward, see bottom row of \Fig{butA}.
This agrees with the magnetic field propagation.
Furthermore, we find some indication of predicted
equatorward migration in the middle of the convection zone, however
this is not as clear as in Run~A1.
We do not find any clear difference between Runs~A2, A3, and A4
(only Run~A3 is shown).
This implies that the size of the coronal envelope is not important
for the magnetic field evolution.
However, in Run~A3t with a modified cooling profile the field is
weaker in the band near the equator and more concentrated at or even
above the surface in comparison with the other runs with coronal
envelopes.
Furthermore, the equatorward migration near the poles is more
pronounced.

Also the migration periods of the large-scale fields change due to the coronal
envelope.
The cycle period of the poleward migrating field in Runs~A2, A3, and
A4 are around two years, which is shorter than in Runs~A1 and
A1c2,
where it is around five years \citep{WKKB14}.
However, the equatorward branch near the poles seems to appear only every
second poleward cycle in Runs~A2, A3, and A4, which give a similar
period as in Runs~A1 and A1c2.
In Run~A3t, the magnetic field near the equator does not show a regular
behavior with a clear cycle.
The equatorward branch near the poles has a period of around two years,
The magnetic field evolution of Run~A5 shows the same features as A3t:
poleward migration or quasi-stationary behavior at low latitudes
and equatorward migration at high latitudes.

In Set~B with slower rotation, the
magnetic field evolution shows a similar dependence on the
cooling profile and the coronal envelope as in Set~A.
In Run~B1 the magnetic field shows equatorward migration,
in particular in the middle of the convection zone, as shown in
\Fig{butA}.
The radius-time diagram is similar to that of Run~A1, but due to the
slower rotation, the cycle period is extended to around ten years.
This is consistent with the predicted dynamo wave scenario; see
bottom row of \Fig{butA}.
Analogously to Run~A1c, the cooling layer in Run~B1c causes the
magnetic field to lose its equatorward migration mode.
The other runs have mostly a stationary mode compared to their rapidly
rotating counterparts, in particular at low latitudes.
As in Set~A, all migration directions of the magnetic field in runs of Set~B agree 
with the predicted propagation.

In general it seems as if the toroidal magnetic field tends to be
strong near the surface in the runs with an extended coronal envelope
(Runs~A1pc, A2, A3, A3t, A4, B3, and B3t), but in the runs with the
vertical field boundary condition (Runs~A1, A1c, A1c2,
B1, and B1c) the boundary condition does not allow this.
Overall, there is good agreement between the predicted and actual direction of
migration of the mean toroidal field.

The magnetic field evolution is similar to that found in \cite{WKMB13}
whose runs have more than two times higher SGS Prandtl
numbers and lower stratification ($\rho/\rho_{\rm surf} =14$), but
the other parameters ($\Rey$, $\Pm$, $\Co$) are comparable with the
runs of this work.
In contrast to this work in the rapid rotating runs of \cite{WKMB13} the field near the
equator seems to continue to propagate equatorward.
In their slower rotating case the field becomes quasi-stationary in
the saturated stage also at high latitudes.
At this point it is not clear if these differences are related to the
larger SGS Prandtl numbers or to the weaker stratification.

\section{Conclusion}

In this work we have studied the influence of the upper boundary on
convectively driven dynamos in spherical wedges.
For this purpose we have added a convectively stable coronal envelope of
different sizes on top of the convectively unstable dynamo region.
This coronal envelope effectively corresponds to having a free boundary
as opposed to a
stress-free radial field boundary condition used in many earlier one-layer
dynamo simulations.
We confirm the result of \cite{WKMB13} that the coronal envelope has
an influence on the dynamo region leading to a change in differential
rotation and magnetic field evolution.
If the radius $\Rc$ of the coronal envelope is just 1\% of the solar
radius $R$, its influence is small and can entirely be related
to small changes in the radial density and temperature profiles.
If the size of the corona is larger ($\Rc\ge 1.2\,R$), the influence
is stronger.
 However, runs with coronal sizes extending higher than $\Rc=1.2\,R$
are nearly identical.

Regarding the hydrothermal properties, the influence of the corona can be
summarized as follows: (i) The radial mass flux across the free surface
does not show any major difference between runs with a small and an
extended coronal envelope.
The radial mass flux in our simulations is too small to have an
influence on the flow properties inside the convection zone. 
(ii) The latitudinal temperature variations due to rotation are
significantly weakened due to the presence of the coronal
envelope.
(iii) This seems to cause the differential rotation profile to become more
spoke-like and weaker in runs with a coronal envelope. 
(iv) This effect can also be seen in the change of the off-diagonal
Reynolds stress components due to the coronal envelope which can be
explained by a change in the $\Lambda$ effect and the anisotropy.

Furthermore, in the cases with $\Ot=5$ and a coronal
envelope, as well as in all runs with $\Ot=3$,
we find the generation of a weak near-surface shear layer at low
latitudes.
We have related this generation to a change of sign in the meridional
Reynolds stress tensor component $Q_{r\theta}$ near the surface in
these runs.
This component contributes to the meridional $\Lambda$ effect and
turns out to be non-zero in all simulations.
Additionally, we have shown that the radial gradients of $Q_{r\theta}$
and $Q_{rr}$ as well as the latitudinal gradient of $Q_{\theta\theta}$ are important near the
surface to balance against differential rotation and the baroclinic
term.
A change of sign in these terms can be associated with the generation
of a near-surface shear layer.
 
The coronal envelope serves as a free top boundary for the
magnetic field.
We find that the dynamo properties are generally strongly
influenced by the choice of boundary conditions.
This can be particularly important when the magnetic field is strong near the
boundary, as was seen recently in connection with the latitudinal boundary
condition for simple $\alpha^2$ mean-field dynamos \citep{CBKK16}
in the wedge geometry that we consider here too.
In all of our simulations
a toroidal field tends to form preferentially near the surface
of the convection zone but it is pushed down by a radial field boundary
condition.
However, with the radial field boundary condition,
the field is only radial over a few grid points below
the boundary; otherwise it is distributed for all simulations
isotropically within the convection zone.
As in \cite{WKKB14}, we compare the migration of the mean
toroidal magnetic field with the predicted propagation direction of
the $\alpha\Omega$ dynamo wave following the Parker--Yoshimura rule.
It turns out that this rule can explain all the different
migration directions found in our simulations.
This is a remarkable result given the variety of the simulations.

We must emphasize that the combination of a dynamo region with
a coronal envelope is still far from realistic.
This is mostly because of the rather
low density contrast and the strong viscous coupling between the two layers.
We must therefore be aware of the possibility of artifacts, for example the
occurrence of a differentially rotating coronal
envelope at higher latitudes may be an example.
This is not normally expected to be the case in the Sun \citep{TKV75}, although
the observations have only been limited to regions where one
sees a rigidly rotating coronal magnetic field.
Indeed, recent work on coronal holes by \cite{LRLM05} claims that
the rigid rotation is only apparent.

Our work may have an impact on our understanding of
the properties of the convection zone and the solar dynamo,
because changing the properties of the boundary and studying its influence
teaches us about the physical properties and dynamical effects
within the solar convection zone.
However, we also have to keep in mind that our model, similarly to other
models of spherical convection for the Sun \cite[e.g.,][]{BMT04,HRY15},
uses an SGS model for the unresolved convective heat flux.
In addition, these models represent the strong radiative cooling
at the surface just by a relaxation term, which ignores the effects
of strong driving of convection by what is known as entropy rain
\citep{Spr97,Bra16}.
This can have dramatic effects on the typical length scales of stellar
convection \citep{CR16}, which in turn would affect the differential
rotation and dynamo properties.

In our simulations, we have confirmed the influence of the latitudinal
temperature distribution on the differential rotation as described by
mean-field models \citep[e.g.,][]{R89} and backed up by simulations
\citep[e.g.,][]{MBT06,WKMB13}.
Only a small temperature difference between pole and equator is needed to cause
the differential rotation to become more spoke-like.
However, if the rotation causes too large latitudinal temperature
variations, the differential rotation becomes dominated by the
Taylor-Proudman theorem and has cylindrical contours.
On the other hand, the strength of the differential rotation benefits
from large latitudinal temperature variations.
Furthermore, our work confirms the results of \cite{HRY15} in that
$Q_{r\theta}$ may be important for the generation of the near-surface
shear layer. 
However, we went a step further and identified the terms in
the thermal wind balance which are important near the surface.
These results appear to be in conflict with those of \cite{KR95,KR05},
which explain the near-surface shear layer solely by the vanishing horizontal
$\Lambda$ effect near the surface. 
We confirm in some simulations that $\LLH$ goes to zero near the surface,
but there we find no clear relation between $\LLH$ and
a negative radial gradient of rotation near the surface.
The associated Reynolds stress component also does not play a role in
the thermal wind balance.
Future helioseismic measurements may yield information about the presence
and distribution of $Q_{r\theta}$ in the Sun.

The fact that the Parker--Yoshimura rule can explain the
migration direction of mean toroidal magnetic field found in our
simulations has an impact on the interpretation of equatorward
migration of the magnetic
field in the Sun.
Applying this rule to the Sun will lead to the generation of toroidal
field in the near-surface shear layer, where negative shear can cause
equatorward migration.
This would also imply that sunspots are formed near the surface by
a local flux concentration mechanism as proposed
by \cite{BKKMR11}, \cite{SN12}, \cite{WLBKR13,WLBKR16}, and
\cite{KBKKR16}.
Furthermore, we find that in our simulations turbulent pumping is
not strong enough to cause a preferred radial orientation of magnetic
field near the surface as in the Sun.
This is possibly due to insufficient stratification in our simulations.
However, a detailed analysis of the effect of turbulent pumping
crucially depends on the test-field method \citep[e.g,][]{SRSRC07,WRKKB16}.

\begin{acknowledgements}
The simulations have been carried out on supercomputers at
GWDG,
on the Max Planck supercomputer at RZG in Garching and in the facilities
hosted by the CSC---IT Center for Science in Espoo, Finland, which are
financed by the Finnish ministry of education.
J.W. acknowledges funding by the Max-Planck/Princeton Center for
Plasma Physics and funding from the People Programme (Marie Curie
Actions) of the European Union's Seventh Framework Programme
(FP7/2007-2013) under REA grant agreement No.\ 623609.
This work was partially supported by the 
Swedish Research Council grants No.\ 621-2011-5076 and 2012-5797 (A.B.),
and the Academy of Finland Centre of Excellence ReSoLVE 272157
(M.J.K., P.J.K. and J.W.) and grants 136189, 140970, 272786 (P.J.K).
\end{acknowledgements}

\bibliographystyle{aa}
\bibliography{paper}

\appendix
\section{Meridional $\Lambda$ effect}
\label{MerLamEffect}

To obtain an expression for the $\Lambda$ effect,
we use the minimal tau approximation \citep{BF02,BF03}; see also the
derivation in \cite{KB08}.
We denote partial time derivatives by a dot and compute
\begin{equation}
\dot{Q}_{ij}=\overline{\dot{u}_i' u_j'}+\overline{u_i' \dot{u}_j'}
\label{MTA}
\end{equation}
in a non-rotating frame of reference
($\meanUU=\meanuu+\eee_\phi\Omega_0 r\sin\theta$), using
\begin{equation}
\dot{u}'_r=2(u'_\theta\overline{U}_\theta+u'_\phi\overline{U}_\phi)/r+...\,,
\label{ur}
\end{equation}
\begin{equation}
\dot{u}'_\theta=-(u'_r\overline{U}_\theta+\overline{U}_r u'_\theta)/r
+2u'_\phi\overline{U}_\phi\cot\theta/r+...\,,
\label{utheta}
\end{equation}
\begin{equation}
\dot{u}'_\phi=-2(u'_r\overline{U}_\phi+\overline{U}_r u'_\phi)/r
-2(u'_\theta\overline{U}_\phi+\overline{U}_\theta u'_\phi)\cot\theta/r+...\,,
\label{uphi}
\end{equation}
where the three dots denote nonlinear and gradient terms of $u_r$,
$u_\theta$, and $u_\phi/r\sin\theta$ that will be neglected.
Below we also consider the case where gradient terms of
$u_\theta$ will be included.
In other words, the base state corresponds to rigid rotation
with a meridional flow $u_\theta\propto r\sin\theta$,
and gradients around this state are neglected.

Inserting \Eqss{ur}{uphi} into \Eq{MTA}, we obtain an expression of the form
\begin{equation}
\dot{Q}_{ij}=L_{ijk}\overline{U_k} + R_{ij},
\label{dotQ}
\end{equation}
where $L_{ijk}$ is a rank 3 tensor (related to the coefficients of
the $\Lambda$ effect) and $R_{ij}$ stands for all remaining terms,
in particular those that stem from the triple correlations.
In the minimal tau approximation we replace those by a relaxation
term with the turbulent correlation time $\tau$, that is
$R_{ij}=-{Q}_{ij}/\tau$.
Inserting this into \Eq{dotQ} and assuming a steady state,
that is $\dot{Q}_{ij}=0$,
using that the background velocity correlation is of the form
$\overline{u_i' u_j'}=\mbox{diag}(\overline{{u'_r}^2},\overline{{u'_\theta}^2},
\overline{{u'_\phi}^2})$, we have
\begin{equation}
{Q}_{r\theta}=2\tau(\overline{{u'_\theta}^2}-\overline{{u'_r}^2})
\,\overline{U}_\theta/r+...\,,
\end{equation}
\begin{equation}
{Q}_{r\phi}=2\tau(\overline{{u'_\phi}^2}-\overline{{u'_r}^2})
\,\Omega\sin\theta+...\,,
\end{equation}
\begin{equation}
{Q}_{\theta\phi}=2\tau(\overline{{u'_\phi}^2}-\overline{{u'_\theta}^2})
\,\Omega\cos\theta+...\,,
\end{equation}
where $\overline{U}_\phi$ has been replaced by $\Omega r\sin\theta$
and gradient terms of $u_\theta/r$ are assumed to vanish, so that
\Eq{utheta} yields
$\dot{u}'_\theta=-2(u'_r\overline{U}_\theta+\overline{U}_r
u'_\theta)/r+...\,$.
We note that, while ${Q}_{r\phi}$ and ${Q}_{\theta\phi}$ are
proportional to $\Omega$, the meridional component
${Q}_{r\theta}$ is proportional to $\overline{U}_\theta$
($\equiv\overline{u}_\theta$).
If we were to allow $u_\theta/r$ to have non-vanishing gradients,
we would have ${Q}_{r\theta}=\tau(2\overline{{u'_\theta}^2}
-\overline{{u'_r}^2})\,\overline{U}_\theta/r+...\,$.
In that case, under isotropic conditions
($\overline{{u'_r}^2}=\overline{{u'_\theta}^2}=\overline{{u'_\phi}^2}$),
$\LLV=\LLH=0$, but $\LLM$ is non-vanishing and would be positive.
Therefore, the non-diffusive contribution to ${Q}_{r\theta}$ would be negative
for a poleward flow.
This is in agreement with the profiles of $\LLM$ and ${Q}_{r\theta}$ in the surface
regions of our simulations; see \Fig{Rey_mer}.

 \section{The anisotropy parameter}
\label{aniso}

The $\Lambda$ effect is related to anisotropy parameters \citep{R80}, defined as
\begin{eqnarray}
A_{\rm M} &=& {\overline{{u'_\theta}^2}-\overline{{u'_r}^2}}\over{\overline{{u'_\theta}^2}+\overline{{u'_r}^2}},\\
A_{\rm V} &=& {\overline{{u'_\phi}^2}-\overline{{u'_r}^2}}\over{\overline{{u'_\phi}^2}+\overline{{u'_r}^2}},\\
A_{\rm H} &=& {\overline{{u'_\phi}^2}-\overline{{u'_\theta}^2}}\over{\overline{{u'_\phi}^2}+\overline{{u'_\theta}^2}}.
\end{eqnarray}

\begin{figure}[t!]
\begin{center}
\includegraphics[width=0.283\columnwidth]{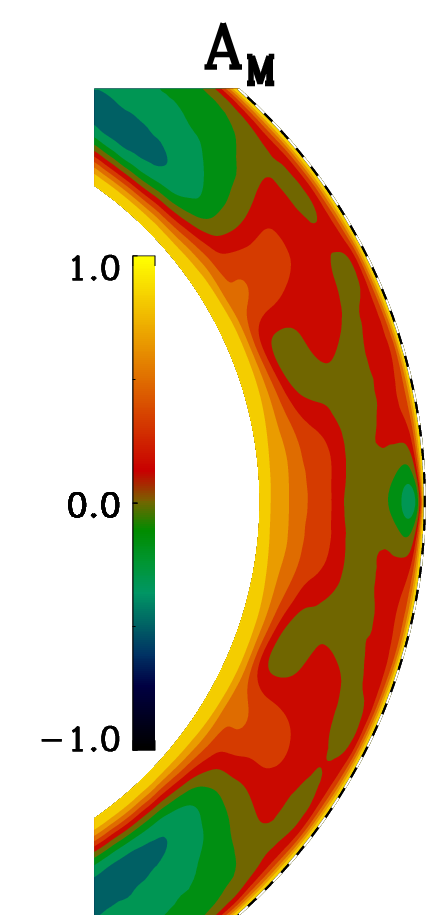}
\includegraphics[width=0.283\columnwidth]{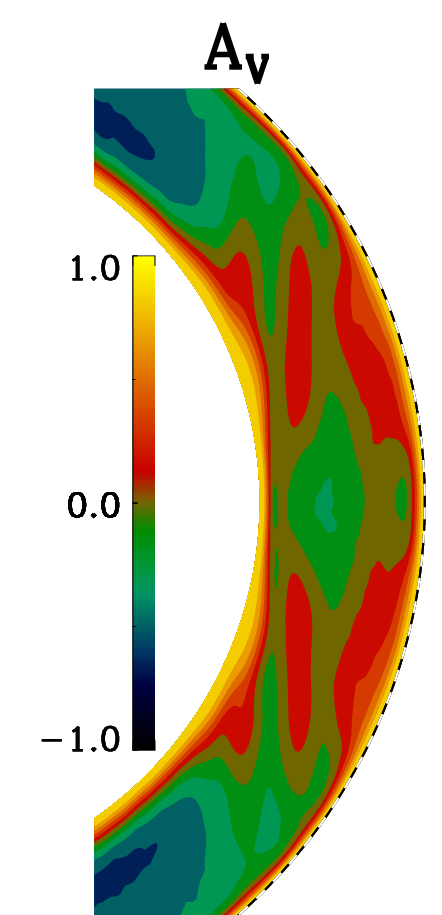}
\includegraphics[width=0.283\columnwidth]{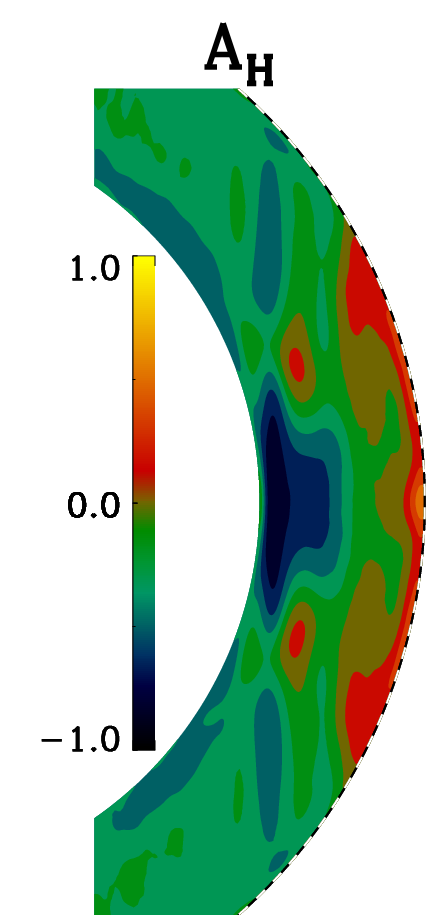}
\includegraphics[width=0.283\columnwidth]{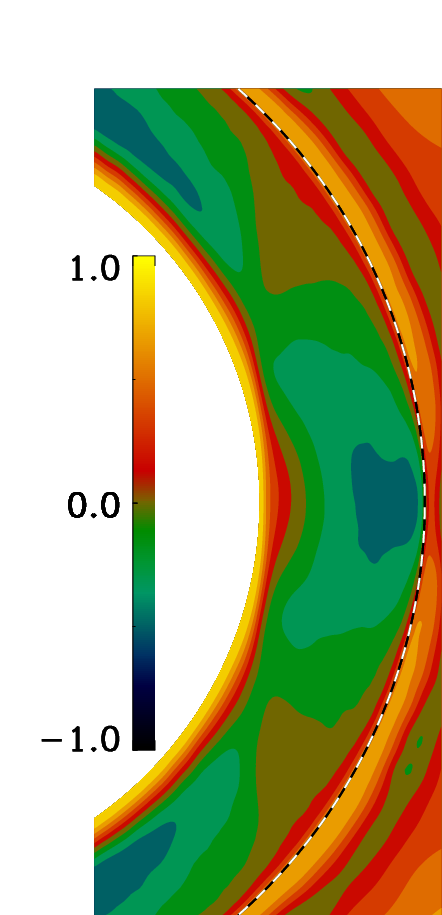}
\includegraphics[width=0.283\columnwidth]{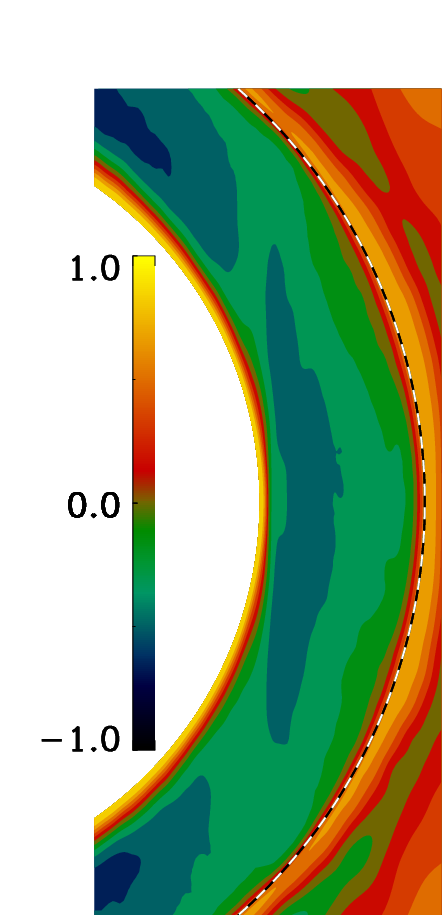}
\includegraphics[width=0.283\columnwidth]{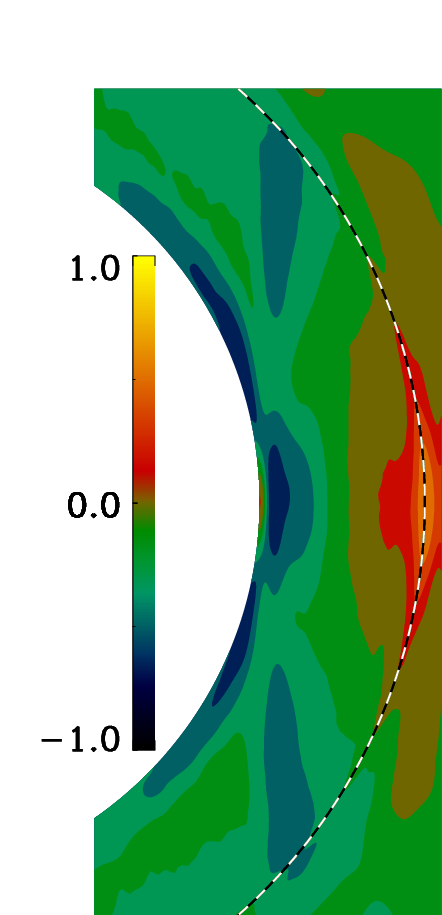}
\end{center}\caption[]{
Anisotropy parameters $A_{\rm M}$, $A_{\rm V}$, $A_{\rm H}$ for Runs~A1
({\it top row}) and A3t ({\it bottom}).
}
\label{An_mer}
\end{figure}

\end{document}